\documentclass[onecolumn]{aastex631}
\usepackage{xcolor} 
\usepackage{multirow} 
\usepackage{longtable} 
\usepackage{booktabs} 
\usepackage{color} 
\newcounter{magicrownumbers} 

\received{08/16/2024}
\revised{...}     
\accepted{08/26/2024}
\submitjournal{\apj}
\shorttitle{T Pyxidis}  
\shortauthors{Godon et al.}
\graphicspath{{./}{figures/}}

\begin{document}


\title{{\bf 
The Accelerating Decline of the Mass Transfer Rate in the Recurrent Nova T Pyxidis   
}
\footnote{Based on observations made with the NASA/ESA Hubble Space Telescope,
obtained from the data archive at the Space Telescope Science Institute.
STScI is operated by the Association of University for Research in 
Astronomy, Inc. under NASA contract NAS 5-26555.}
}

\correspondingauthor{Patrick Godon} 
\email{patrick.godon@villanova.edu} 

\author[0000-0002-4806-5319]{Patrick Godon}  
\affiliation{Department of Physics and Planetary Science, Villanova University, Villanova, PA 19085,  USA} 

\author[0000-0003-4440-0551]{Edward M. Sion}
\affiliation{Department of Physics and Planetary Science, Villanova University, Villanova, PA 19085,  USA}

\author[0000-0002-3742-8460]{Robert E. Williams} 
\affiliation{Space Telescope Science Institute, Baltimore, MD 21218, USA} 

\author[0000-0003-0156-3377]{Matthew J. Darnley}
\affiliation{Astrophysics Research Institute, Liverpool John Moores University, IC2 Liverpool
Science Park, Liverpool, L3 5RF, UK} 

\author[0000-0002-8286-8094]{Jennifer L. Sokoloski}
\affiliation{Columbia Astrophysics Laboratory and Department of Physics, Columbia University, New York, NY 10027, USA}

\author[0000-0002-7491-7052]{Stephen S. Lawrence}
\affiliation{Department of Physics and Astronomy,Hofstra University, Hamstead, NY 11549, USA} 

\begin{abstract}

The recurrent nova T Pyxidis has erupted six times since 1890, with its last outburst in 2011,
and the relatively short recurrence time between classical nova explosions indicates 
that T Pyx must have a massive white dwarf accreting at a high rate. 
It is believed that, since its outburst in 1890, the mass transfer rate in T Pyx was very large 
due to a feedback loop where the secondary is heated by the hot white dwarf.  
The feedback loop has been slowly shutting off, reducing the mass transfer rate,
and thereby explaining the magnitude decline of T Pyx from $\sim13.8$ (before 1890)
to 15.7 just before the 2011 eruption.  
We present an analysis of the latest {\it Hubble Space Telescope} (HST) far ultraviolet and optical spectra,
obtained 12 years after the 2011 outburst, showing that the mass transfer rate has 
been steadily declining and is now below its pre-outburst level by about 40\%:   
$\dot{M} \sim 1-3\times 10^{-7}M_\odot$/yr 
for a WD mass of $\sim 1.0-1.4 M_\odot$, an inclination of 
$50^\circ - 60^\circ$, 
reddening $E(B-V)=0.30 \pm 0.05$ and a Gaia DR3 distance of 
$2860^{+816}_{-471}$~pc.  
This steady decrease in the mass transfer rate in the $\sim$decade after the 
2011 ourbutst is in sharp contrast with the more constant pre-outburst 
UV continuum flux level from archival international ultraviolet explorer (IUE) spectra. 
The flux (i.e. $\dot{M}$) decline rate is 29 times faster now in the last $\sim$decade 
than observed since 1890 to $\sim$2010.   
The feedback loop shut off seems to be accelerating, at least in the decade following
its 2011 outburst. In all eventualities, our analysis confirms that T Pyx is going through
an unusually peculiar short-lived phase.  

\end{abstract}


\section{{\bf Introduction}}

Cataclysmic Variables (CVs) are short period interacting binaries
where a white dwarf star (WD) accretes matter from its companion star
(the donor) filling its Roche lobe. The transfer of material can 
be continue (as for UX UMa novalikes), sporadic (as for VY Scl novalikes
and some dwarf nova systems), or almost periodic (as for many dwarf novae)
and translates into a change in luminosity on time scales of days
to months or even years \citep[e.g.][]{hac93,lad94}. 
Over time (years to millennia), 
the accreting WDs in CVs accumulate a layer of hydrogen-rich 
material which, when the layer has reached a critical mass,  
provides enough temperature and 
pressure at its base to initiate a thermonuclear (TNR) 
runaway: the {\it classical nova} explosion \citep[or simply nova;][]{sch49,pac65,sta72}.  
The larger the WD mass and the higher the mass accretion rate onto it, 
the shorter the recurring time between such TNR nova explosions
\citep[e.g.][see also the remarkable recurrent nova M31N 2008-12a,
with annual eruptions \citep{dar17}]{yar05}. 
CVs that have suffered a classical nova explosion 
are called novae, and those that have experienced more than one
nova explosion are referred to as recurrent novae (RNe;
for a review on classical novae - see \citet{bod08}). 
While mass accumulates onto the white dwarf during quiescence between recurring 
nova explosions, mass is also ejected during the nova explosions themselves, and the question whether the 
white dwarf mass increases or decreases over its life-time is still a matter of debate
\citep{sta20,hil20,hil21}.   
As a consequence, recurrent novae  are potential progenitors
of Type Ia Supernovae (SNe Ia) as their WD may grow in mass and 
reach the Chandrasekhar limit for a supernova explosion
\citep[][; see \citet{liv11} for a review]{whe73}.   
As such, accreting WDs in CVs are the site of   
some of the most violent eruptions in the Galaxy, exhibiting large 
luminosity changes on time-scales of $\sim$days to millennia.  

T Pyxidis is a CV that had six nova eruptions since 1890: 
in 1902, 1920, 1944, 1967, with the last outburst in 2011
\citep[the first eruption being in 1890; ][]{sch13}.   
Because of that, T Pyx is one of the most-studied RNe,
it has also become one of the most enigmatic RNe,
and certainly the most famous RN in the Milky Way. 
T Pyx is one of the three known short orbital period RNe 
(together with IM Nor and CI Aql), 
it is the only RN with a nova shell \citep{due79} 
and its rise to outburst is characterized as slow
\citep{sch10}. 
The expansion of the shell is believed to have originated from a normal 
classical nova eruption around the year 1866 \citep{sch10}.

Its relatively short (and {\it increasing}) recurrence time 
(12, 18, 24, 23, and 44 years) indicates, on theoretical ground
\citep{sta85} that its WD must be massive
accreting at a high rate 
\citep[possibly $\dot{M}\sim 1-2\times 10^{-7}M_\odot$/yr;][]{yar05}. 
And indeed, optical and ultraviolet (UV) analyses \citep[e.g.][]{sel08,pat17,god18} derived 
a mass transfer rate (disk luminosity) anywhere between $10^{-6}$ and $10^{-8}M_\odot$/yr
(depending on the assumed WD mass, distance, reddening, and inclination).   
However, with an orbital period of 1.83~hr, the mass transfer
rate (due to angular momentum loss by gravitational radiation) 
should be very low, of the order $2 \times 10^{-11}M_\odot$/yr,  
as it is the case for CV systems with an orbital period
below 2~hr \citep{pat84}.   
In order to explain the mass transfer/accretion\footnote{
Note that we are neglecting here outflow from the disk and WD and
use the term `mass transfer rate' when considering the disk
(or the Roche lobe overflow of the secondary), and `mass accretion  
rate' when considering the accretion disk and WD, assuming that they are nearly equal:
we use $\dot{M}$ for both. It is understood that the mass accretion
rate might be slightly smaller than the mass transfer rate due to 
possible outflow. 
} rate discrepancy of T Pyx and other
novae, several theories have been advanced. 

\citet{shara86} suggested that novae {\it hibernate} for millennia between
eruptions to explain their (very low) space density in the
solar neighborhood and justify the fact that old novae have
low $\dot{M}$ while recent novae have a higher mass accretion. 
During a nova eruption, mass loss dominates, increasing the binary
separation and Roche lobe radius. As a consequence, the secondary loses contact   
with the inflated Roche lobe and mass transfer basically stops after the eruption and 
after irradiation from the cooling WD becomes negligible. This explains the
high $\dot{M}$ after the eruption and its decline thereafter, up to the point
where hibernation starts ($\dot{M} < 10^{-12}M_\odot$/yr), lasting 1000s of years,
during which the binary separation decreases slowly due to angular momentum loss from 
magnetic breaking (above the gap) or gravitational radiation (below the gap). 
\citet{shara86} suggested that in this manner 
most novae spend 90-99\% of their lives as detached binary.  
In this scenario, the high mass transfer rate would have been sustained    
by the irradiation of the secondary by the white dwarf, itself heated due to  
accretion \citep{kni00}.  
Such a self-sustained feedback loop process would have been triggered 
during a classical nova eruption in 1866 \citep{sch10}, where the high mass
accretion rate would occur with nuclear burning on the WD surface 
(self-sustained supersoft source).  
However, it has been shown \citep[see ][Fig.1]{sch13},  
that the B-magnitude of T Pyx has been steadily decreasing 
from $B=13.8$ before the 1890 eruption to $B=15.7$ just before the 2011 eruption,
indicating that the self-sustained feedback loop between the WD
and secondary might be shutting off, in agreement with the hibernation theory.   

It has also been proposed \citep{kni22} that the high mass transfer rate in T Pyx 
could be the result of the evolution of triple star system, where the inner binary (WD + donor star) 
would become so eccentric that mass transfer is triggered at periastron, driving 
the secondary out of thermal equilibrium.  

\citep{pat17} showed that with a mass transfer rate of $\sim 10^{-7}M_\odot$/yr and a nova ejecta mass 
of $3 \times 10^{-5}M_\odot$ (6.7 times larger than the accreted mass between novae),  
the present series of nova eruptions are eroding the WD, and the secondary 
will evaporate in $10^5$~yr, unless the recurrent nova eruptions are short-lived.    

All these analyses agree that T Pyx must be going through a very unusual 
and short-lived phase in its life 
\citep[according to ][possibly its last phase]{pat17}. 

In the current work, we present an analysis of the latest  
Hubble Space Telescope (HST) UV and optical spectra from         
March 2023. The UV spectrum was obtained with 
the Cosmic Origin Spectrograph (COS), while the optical spectrum was 
obtained using the Space Telescope Imaging Spectrograph (STIS).     
This is the first combined optical (STIS) and FUV (COS) 
spectroscopic observation of T Pyx during the deep quiescent phase
to model the accretion disk: the inner disk radiates mainly in the UV,  
while the outer disk radiates mainly in the optical. 
The results of our analysis indicate that the mass accretion rate is still decreasing 
compared to the HST data from 2015-2016 \citep{god18} and 2012-2013 \citep{god14},  
it has now reached a level that is 40\% below its pre-outburst IUE value.  
Such a steady decrease in $\dot{M}$ is unexpected, since 
all the IUE spectra obtained through
the 90's have the same flux level as the 1980 IUE spectrum
and show no drop in flux (except for orbital variation). 
This could indicate that the decrease in the mass transfer rate started to 
accelerate after the 2011 outburst.   
 
In the next section we discuss the system parameters that we adopted in the
present work; in \S 3 we present the latest HST data together with archival 
data for our analysis; the tools we used and the results obtained are 
presented in \S 4, follow by a discussion and summary in the last section.

\section{{\bf System Parameters}}

In our previous analysis of T Pyx \citep{god18} we analyzed HST COS UV spectra 
obtained in October 2015 and June 2016 and investigated the effect of the assumed
WD mass ($0.7 M_\odot \le  M_{\rm wd}\le 1.35 M_\odot$), reddening ($0.25 \le E(B-V) \le 0.50$), 
distance ($2.8~kpc \le d \le 4.8~kpc$), and inclination ($20^\circ \le i \le 60^\circ$) 
on the results ($\dot{M}$). Therefore, we will not repeat this in the current work. 
Instead, and unless otherwise indicated, we assume here a large WD mass ($M_{\rm wd} = 1.00-1.37 M_\odot$), 
a reddening of $E(B-V)=0.30\pm0.05$, 
a Gaia DR3 parallax-derived distance of 2860$^{+816}_{-471}$~pc, 
and an inclination $i=50^\circ-60^\circ$. Here below we justify our choice.
The value of the system parameters we use for the analysis are listed in Table 1. 

Taking the latest DR3 Gaia parallax to the system and following \citet{sch18}, we compute a distance 
of $2860^{+816}_{-471}$~pc, which is smaller than the distance originally 
derived from the light echo \citep[4.8$\pm$0.5~pc][]{sok13} and  
the distance we used in our previous spectral analysis based on the DR2 Gaia parallax 
\citep[$3277^{+521}_{-395}$~pc][]{god18}.  

Using recent data from the Multi Unit Spectroscopic Explorer (MUSE)
from the European Southern Observatory (ESO) in Chile, \citet{izz24} characterize 
the morphology of the ejecta surrounding the system. 
They found that the expelled material consists of a ring of matter together with a 
bipolar outflow perpendicular to the ring. 
The inclination of the remnant along the line of sight is $i=63.7^\circ$, 
and is expanding at a velocity of $472^{+77}_{-72}$km/s. 
They put an upper limit
to the bipolar outflow ejecta mass, $M_{\rm ej,b} < (3 \pm 1) \times 10^{-6}M_\odot$,   
which is lower than previous estimates. 
It is believed that the bipolar outflow originated from the 2011 outburst
(since it wasn't observed before, and was first observed by HST in 2014). 
Consequently, we consider here an inclination $i\approx 50-60^\circ$,  
\citep[as suggested by][]{pat17} to account for the large amplitude ($\sim 20$\%) optical and
UV modulation in the continuum flux level as a function of the orbital phase and
to agree with the analysis of \citet{izz24}.

As to the WD mass, on the one hand, based on the short recurrence time of T Pyx outbursts (of the order of 20~yr or so), 
the theory predicts \citep[e.g.][]{sta85,web87,sch10} that 
the WD in T Pyx must be very massive
\citep[possibly near-Chandrasekhar: $1.37 M_\odot$][]{sel08}  
accreting at a very large rate.
On the other hand, X-ray observational evidence tends to point to a lower 
mass of the order of $1.00 - 1.15 M_\odot$ \citep[e.g.][based on X-ray observations in the months/year following
the 2011 outburst]{tof13,cho14}. 
Accordingly, in the current analysis we assume a WD mass 
$M_{\rm wd}=1.0$, $1.2$, and $1.37 \times M_\odot$, 
and we disregard the low WD mass ($0.7 M_\odot$) derived by \citet{uth10}, since it was retracted \citep{kni19}.  
This is in line with \citet{sha18} who showed that 
extensive simulations of nova eruptions combined together with observational
databases of outburst characteristics of Galactic classical novae and recurrent novae 
yield for T Pyx a WD mass of $1.23 M_\odot$ ($\pm 0.1 M_\odot$ or so) with a mass accretion rate of 
$6.3 \times 10^{-8}M_\odot$/yr (but no error estimated given on $\dot{M}$) 
for the 44 year inter-outburst period between 1967 and 2011.    

For the reddening we limit ourselves to the value we derived previously in \citet{god18}.  

We must stress that the uncertainties in the values of the system parameters
(WD mass, distance, inclination, extinction, chemical abundances, etc.. ;
which are used a input for the analysis) are {\it relatively} 
much larger than the errors in the analysis results that depend on them. 

\clearpage 

\begin{small} 
\begin{deluxetable}{llcclllcc}[h!] 
\tablewidth{0pt}
\tablecaption{T Pyxidis System Parameters
\label{syspar} 
} 
\tablehead{ 
Parameter           &~~$P_{\rm orb}$& $i$      & $\Pi$ - Gaia         & ~~~~$d$            & $E(B-V)$       & $M_{\rm wd}$   \\       
Units               & ~(hr)       &  (deg)     & (mas)                & ~~(pc)             &                & ($M_{\odot}$)              
}
\startdata
Adopted Value       & 1.8295      &$50-60$    &$0.34674\pm0.0287$  & $2860^{+816}_{-471}$ & $0.30\pm0.05$ & $1.0, 1.2, 1.37$            \\[1pt]
\enddata
\tablecomments{
Unless otherwise specified, these are the values of the system parameters we used
in the present analysis of the 2023 HST spectra (see text for details). 
} 
\end{deluxetable} 
\end{small}

\section{{\bf The Data}} 

In this research we analyze the most recent HST UV-Optical spectral data
we obtained in 2023. For comparison and to complement the analysis we also
present the HST UV data we obtained in 2018-2019, HST UV data from our previous analyses
(2012, 2013, 2015, 2016), IUE pre-outburst data, 
and some never-published HST optical data obtained in 2014. 
Since the IUE data and our previous HST UV data were already presented in  
\citet{god18}, we tabulate here only the data that weren't presented elsewhere:
COS UV data from Oct 2018, Feb 2019, March 2023, STIS optical data
from March 2023, and STIS optical data obtained in 2014 (PI A. Crotts)
which were never published. All the data are listed in Table 2. 

These observations were obtained with four different instrument configurations  
as follows. 
\\ 
\indent 
-1) The COS instrument (FUV MAMA, TIME-TAG mode) was set up with the PSA aperture, with the G130M grating
with a central wavelength of 1055~\AA , producing a spectrum starting at  925~\AA\ 
all the way to 1200~\AA , with a small gap near 1050~\AA (and therefore covering all
the series of the hydrogen Lyman transitions, except Ly$\alpha$).  
\\ 
\indent 
-2) The COS instrument (FUV MAMA, TIME-TAG mode) was set up with the PSA aperture, with the G140L grating
with a central wavelength of 1105~\AA , producing a spectrum from 1100~\AA\ 
to $\sim$2100~\AA , covering the H Ly$\alpha$ absorption feature.   
\\
\indent 
-3) The STIS instrument (CCD, ACCUM mode) was set up with the G430L grating centered at 4300~\AA ,
generating a spectrum from $\sim3,000$~\AA\ to $\sim$5,700~\AA .
\\
\indent 
-4) The STIS instrument (CCD, ACCUM mode) was set up with the G750L grating centered at 7751~\AA ,
generating a spectrum from $\sim5,250$~\AA\ to almost 10,000~\AA , thereby covering 
the optical and near infrared region.   

The COS data were processed with CALCOS version 3.4.4 and the
STIS data were processed with CALSTIS version 3.4.2. 
We used the x1d and sx1 files to extract the 1D spectra from each individual
exposures, and used the x1dsum files to extract spectra from 
co-added exposures (such as for the COS data obtained on 4 different
positions of the detector). 

\subsection{\bf The 2023 HST COS FUV and STIS Optical Data.}  

\begin{figure}
\includegraphics[scale=0.55,trim=-4cm 0 0 0]{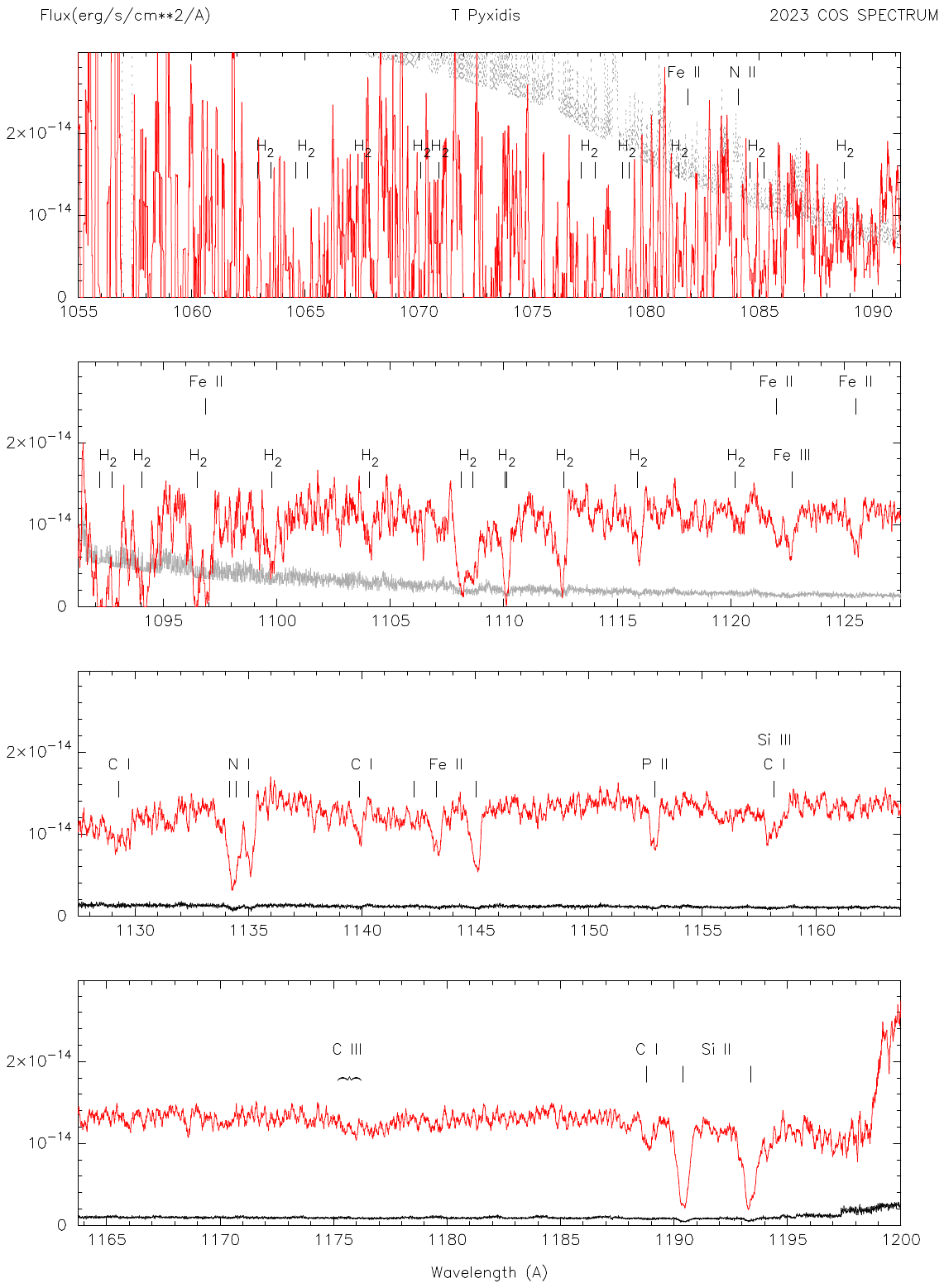} 
\caption{The 2023 HST COS G130M (1055~\AA ) spectrum of T Pyx with line identifications. 
The spectrum is in red; for convenience and clarity the error is in dashed grey in the upper/first 
panel, grey in the second panel, and black in the two lower panels. 
The spectrum has not been dereddened. 
\label{fuvslines}
}
\end{figure}

\begin{figure}
\includegraphics[scale=0.55,trim=-4cm 0 0 0]{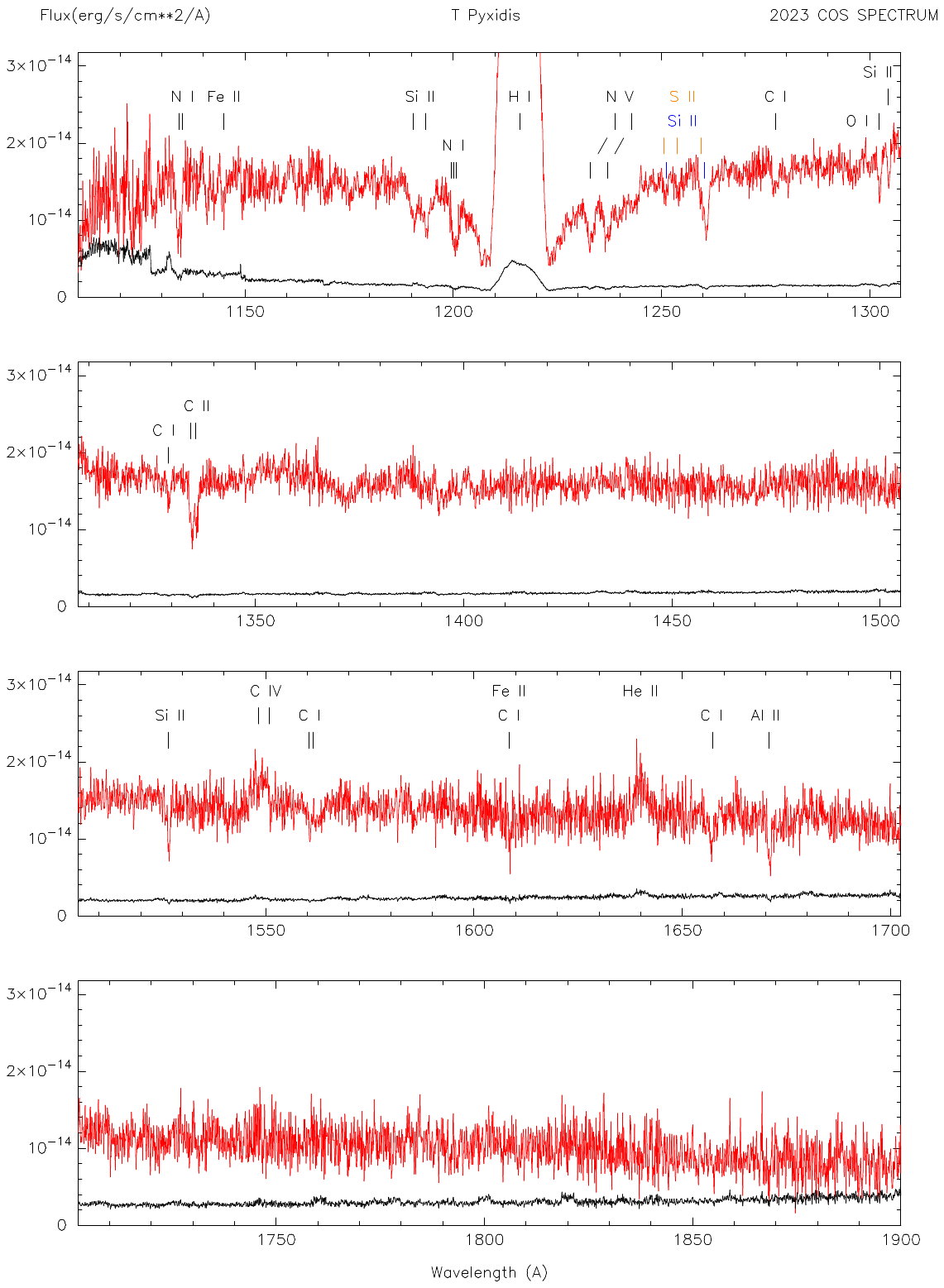} 
\caption{The 2023 HST COS G140L (1105~\AA ) spectrum of T Pyx with line identifications. 
The spectrum is in red, the error in black. 
The S\,{\sc i} and Si\,{\sc ii} lines near 1250-1260~\AA\ are in color
so that they can be identified separately. Note the strongly blue-shifted N\,{\sc v} (1240~\AA) doublet.  
The spectrum has not been dereddened. 
\label{fuvllines}
}
\end{figure}

\begin{figure}[b!]  
\includegraphics[scale=0.44,trim=-5cm 0 0 0]{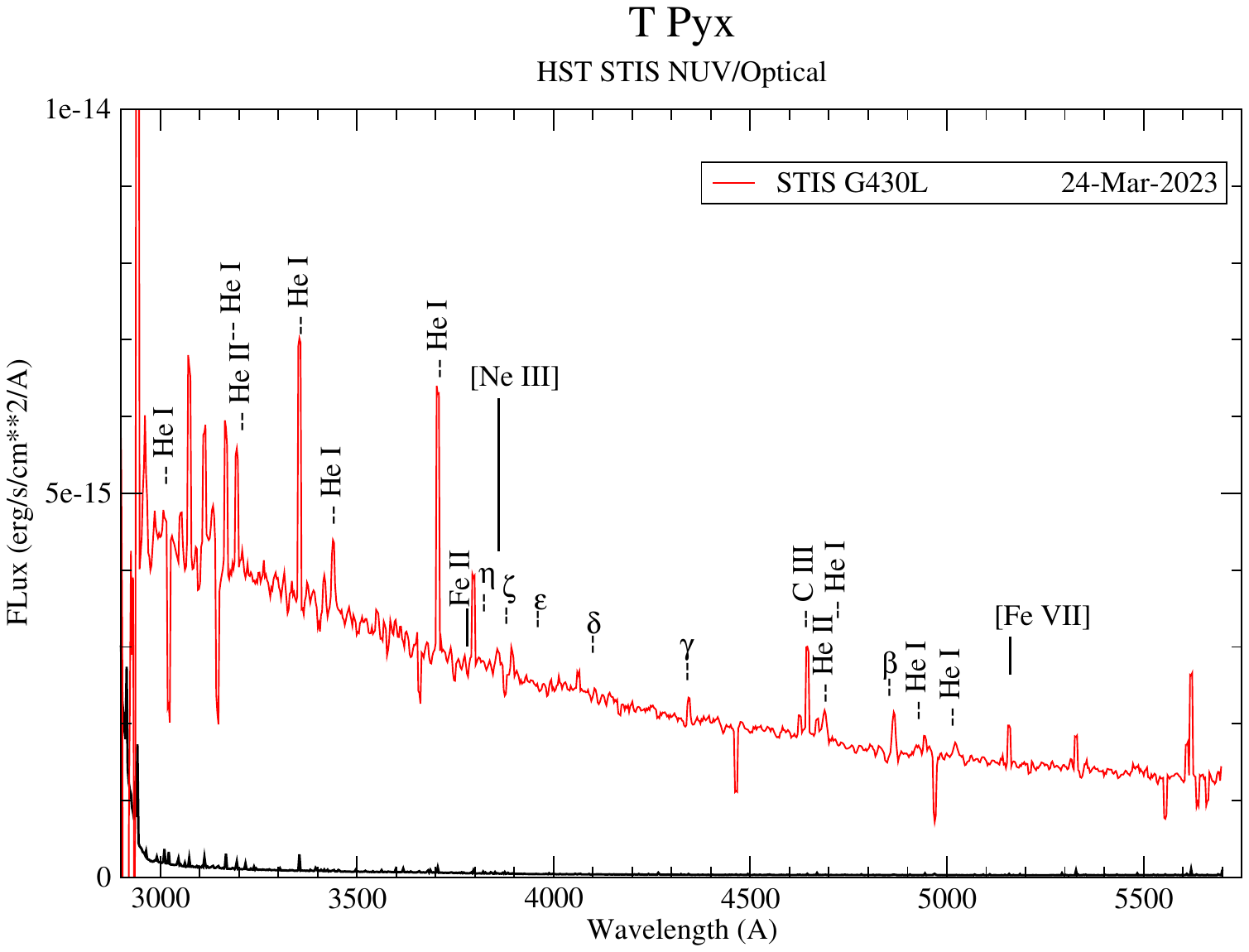} 
\vspace{-0.5cm} 
\caption{ 
The 2023 HST STIS G430L (4300~\AA ) spectrum (in red) with line 
identifications; the error is in black.  
This spectrum has not been corrected for extinction.  
\label{opt1}
}
\includegraphics[scale=0.44,trim=-5cm 0 0 0]{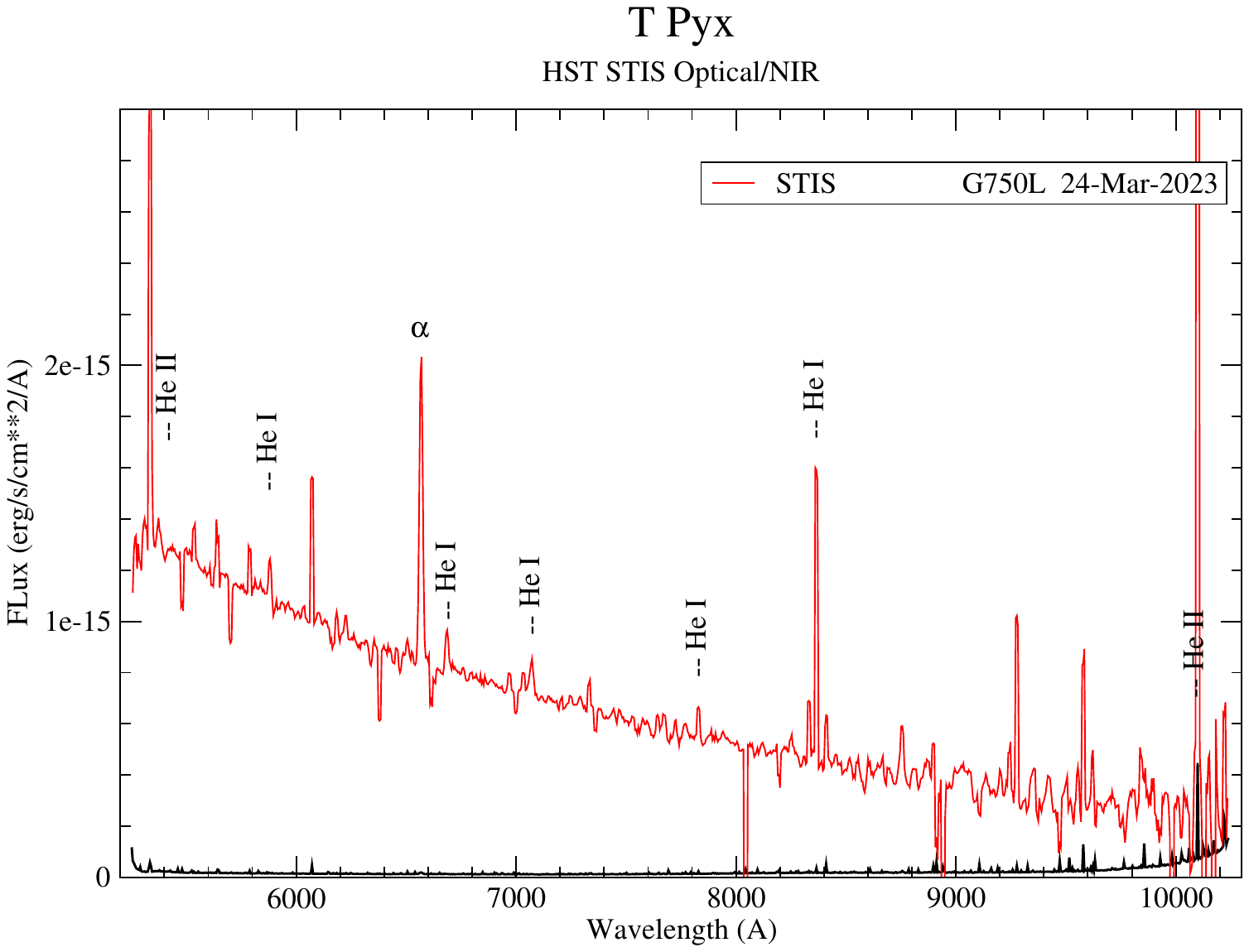} 
\vspace{-0.5cm} 
\caption{ 
The 2023 HST STIS G750L (7751~\AA ) spectrum (in red) 
with line identifications (the error is in black).  
The spectrum exhibits mostly hydrogen and helium emission lines. 
This spectrum has not been corrected for extinction.  
\label{opt2}
}
\end{figure}

The 2023 data consist of one of each instrument configuration above and were all obtained concurrently,
the same day, March 24th, 2023, between about midnight to 11am - see Table 2. 
Namely the 2023 data cover the FUV, UV, optical, and NIR, and produce the only concurrent UV-optical-NIR  
spectra of T Pyx from $\sim900$~\AA\ to $\sim$10,000~\AA\ (with a gap between 2000~\AA\ and 3000~\AA ). 
These 4 concurrent UV-optical spectra are of special importance, since they are the only ones obtained 
concurrently after the 2011 outburst and during {\it deep} quiescence when all the emission is from 
the accretion disk; These 4 2023 spectra are the focus of the present analysis and are 
modelled in \S 4 with an accretion disk.  
We present these four spectra in Figs.\ref{fuvslines}, \ref{fuvllines}, \ref{opt1},
and \ref{opt2}, in order of increasing wavelength.

\begin{deluxetable}{cccccccccc}[h!] 
\tablewidth{0pt}
\tablecaption{Observation Log 
\label{obslog} 
} 
\tablehead{ 
 Instrument  & Apertures & Filter   & Central     & Date       & Time     & ExpTime   &  Data   & MODE &  Project       \\       
             &           & Gratings & $\lambda$(\AA)& yyyy-mm-dd & hh:mm:ss & (s)     &   ID    &      &    ID       
}
\startdata
   STIS     & 52x0.1   & G430L     & 4300        & 2023-03-24 & 08:18:47 & 1699    & OEWH02010  & ACCUM  & 17190  \\ 
   STIS     & 52x0.1   & G750L     & 7751        & 2023-03-24 & 10:26:35 & 2340    & OEWH02020  & ACCUM  & 17190  \\ 
   COS      & PSA      & G140L     & 1105        & 2023-03-24 & 00:19:41 & 1912    & LEWH01010  & TIME-TAG  & 17190  \\ 
   COS      & PSA      & G130M     & 1055        & 2023-03-24 & 01:46:13 & 2405    & LEWH01020  & TIME-TAG  & 17190 \\ 
   COS      & PSA      & G130M     & 1055        & 2019-02-01 & 12:58:11 & 1836    & LDG002010  & TIME-TAG  & 15184  \\ 
   COS      & PSA      & G140L     & 1105        & 2018-10-04 & 00:30:11 & 1876    & LDG001010  & TIME-TAG  & 15184  \\ 
   STIS     & 52x2     & G430L     & 4300        & 2014-07-21 & 02:29:59 &  378    & OCIQ02010  & ACCUM  & 13796  \\ 
            &          &           &             &            & 02:38:11 &  378    & OCIQ02020  & ACCUM  & 13796   \\ 
            &          &           &             &            & 02:46:23 &  378    & OCIQ02030  & ACCUM  & 13796   \\ 
            &          &           &             &            & 02:54:35 &  378    & OCIQ02040  & ACCUM  & 13796   \\ 
            &          &           &             &            & 03:51:04 &  558    & OCIQ02050  & ACCUM  & 13796   \\ 
            &          &           &             &            & 04:04:42 &  558    & OCIQ02060  & ACCUM  & 13796   \\ 
            &          &           &             &            & 04:15:54 &  558    & OCIQ02070  & ACCUM  & 13796   \\ 
            &          &           &             &            & 04:27:06 &  558    & OCIQ02080  & ACCUM  & 13796   \\ 
            &          &           &             &            & 05:26:37 &  552    & OCIQ02090  & ACCUM  & 13796   \\ 
            &          &           &             &            & 05:40:35 &  552    & OCIQ020A0  & ACCUM  & 13796   \\ 
            &          &           &             &            & 05:51:41 &  552    & OCIQ020B0  & ACCUM  & 13796   \\ 
            &          &           &             &            & 06:02:47 &  552    & OCIQ020C0  & ACCUM  & 13796   \\ 
            &          &           &             &            & 07:08:57 &  544    & OCIQ020D0  & ACCUM  & 13796   \\ 
            &          &           &             &            & 07:19:55 &  544    & OCIQ020E0  & ACCUM  & 13796   \\ 
            &          &           &             &            & 07:30:53 &  544    & OCIQ020F0  & ACCUM  & 13796   \\ 
            &          &           &             &            & 08:40:09 &  544    & OCIQ020G0  & ACCUM  & 13796   \\ 
   STIS     & 52x2     & G750L     & 7751        & 2014-07-23 & 21:23:16 &  373    & OCIQ03010  & ACCUM  & 13796   \\ 
            &          &           &             &            & 21:31:24 &  373    & OCIQ03020  & ACCUM  & 13796   \\ 
            &          &           &             &            & 21:39:32 &  373    & OCIQ03030  & ACCUM  & 13796   \\ 
            &          &           &             &            & 21:47:40 &  373    & OCIQ03040  & ACCUM  & 13796   \\ 
            &          &           &             &            & 22:44:01 &  558    & OCIQ03050  & ACCUM  & 13796   \\ 
            &          &           &             &            & 22:57:39 &  558    & OCIQ03060  & ACCUM  & 13796   \\ 
            &          &           &             &            & 23:08:51 &  558    & OCIQ03070  & ACCUM  & 13796   \\ 
            &          &           &             &            & 23:20:03 &  558    & OCIQ03080  & ACCUM  & 13796   \\ 
            &          &           &             & 2014-07-24 & 00:19:33 &  552    & OCIQ03090  & ACCUM  & 13796   \\ 
            &          &           &             &            & 00:33:31 &  552    & OCIQ030A0  & ACCUM  & 13796   \\ 
            &          &           &             &            & 00:44:37 &  552    & OCIQ030B0  & ACCUM  & 13796   \\ 
            &          &           &             &            & 00:55:43 &  552    & OCIQ030C0  & ACCUM  & 13796   \\ 
            &          &           &             &            & 01:55:53 &  543    & OCIQ030D0  & ACCUM  & 13796   \\ 
            &          &           &             &            & 02:06:51 &  543    & OCIQ030E0  & ACCUM  & 13796   \\ 
            &          &           &             &            & 02:17:49 &  543    & OCIQ030F0  & ACCUM  & 13796   \\ 
            &          &           &             &            & 02:28:47 &  543    & OCIQ030G0  & ACCUM  & 13796   \\ 
\enddata
\tablecomments{
The time (hh:mm:ss) is the start time for each exposure. 
All the data presented were obtained from the Mikulski Archive for Space Telescope (MAST) at the
Space Telescope Science Institute, Baltimore, MD, USA. The specific spectral data listed
above can be accessed via 
\dataset[10.17909/2d6h-qy95]{https://doi.org/10.17909/2d6h-qy95}. 
COS Data were processed through the pipelines with CALCOS version 3.4.4.
STIS Data were processed through the pipelines with CALSTIS version 3.4.2.
} 
\end{deluxetable}

\clearpage 

The short wavelength COS (FUV) spectrum is displayed in Fig.\ref{fuvslines} on four
panels, it is very noisy below 1090~\AA\ (first/upper panel).  
All the absorption lines are from the interstellar medium (ISM), dominated mainly by molecular hydrogen
(H$_2$), with some C\,{\sc i}, and Fe\,{\sc ii} lines. The identification of the H$_2$ molecular lines
by their band, upper vibrational level, and rotational transition can be found e.g. in \citet{sem01}. 
The rather flat shape of the continuum flux level indicates that the emitting source is rather hot 
and is consistent with the inner part of the accretion disk.  

\clearpage 

The long wavelength COS (UV) spectrum is displayed in Fig.\ref{fuvllines},
also on four panels.   
Except for the N\,{\sc v} doublet (which is blue shifted by $\sim6$~\AA ), 
all the absorption lines are from the ISM. Due to the relatively lower continuum
flux level during deep quiescence, the S/N is not high enough to detect all the
ISM lines which were observed in the early phase following the outburst 
by \citet{deg14}.

Each COS spectrum is generated from the sum of 4 subexposures, each obtained
on a different location on the detector. 
We checked the 4 subexposures each of the two 2023 COS spectra and did not
find, within the amplitude of the noise/error, any variation in the width, depth, and 
wavelength of the absorption lines that could reveal orbital modulation,
even for the N\,{\sc v} doublet. 
However, this is likely an indication that the  
subexposures are too noisy to extract any significant information.
The rest wavelength of the N\,{\sc v} doublet lines are 1238.821~\AA\ \& 1242.804~\AA\
\citep{kra23}, 
and to within $\pm$0.1\AA\ the observed wavelengths in the 4 subexposures 
are    1232.9~\AA\ \& 1237.0~\AA , 1233.3~\AA\ \& 1237.1~\AA ,   
1233.1~\AA\ \& 1237.0~\AA , and 1233.0~\AA\ \& 1237.1~\AA .  
This gives an average blue shift of $\sim$5.7$\pm$0.2 \AA , 
which at 1240~\AA\ corresponds to a velocity of 1,384$\pm$49~km/s.

\clearpage 

The STIS optical-NIR spectra are presented in Figs.\ref{opt1} \& \ref{opt2}.
Contrary to the previous optical spectra, these 2 spectra are no dominated
by nebular emission,  
they exhibit absorption and emission lines from hydrogen and helium,
and we tentatively identify some weak emission lines from 
C\,{\sc iii} (4650~\AA)    and 
[Fe\,{\sc vii}] (5168~\AA).

\clearpage 

\subsection{\bf Earlier Archival UV and Optical Data}  

The existing pre-outburst UV archival data of T Pyxidis 
consist of more than 50 IUE SWP+LWP spectra from 1980 
($\sim$13 years after the Dec 1966/Jan 1967 outburst) 
through the 90's and one Galex (FUV + NUV) spectrum taken at
the end of 2005. The pre-outburst data reveal a UV continuum flux level remarkably 
constant, except for an orbital phase modulation.    

The HST COS \& STIS UV spectra, all obtained post-outburst,  
follow the decline of the system into its quiescent state, 
starting May 2011. By December 2012, the strong broad emission lines 
had disappeared and the UV continuum flux level
had reached its pre-outburst (IUE) level, after what  
the UV flux continued to decrease, but more slowly \citep[see][for a review]{god18}. 

\begin{figure}[b!] 
\includegraphics[scale=0.39,trim=1.5cm 0 0 0]{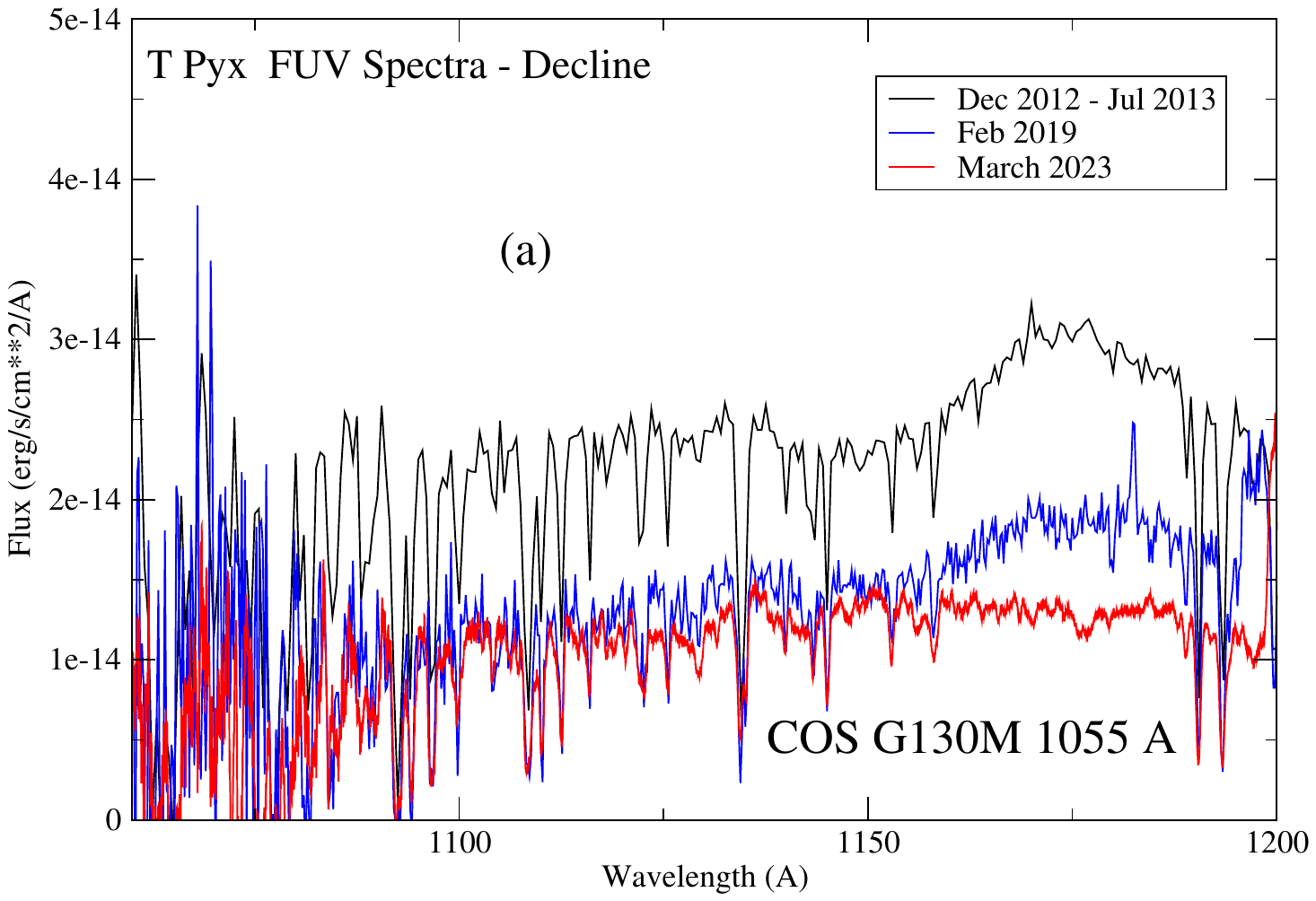}
\includegraphics[scale=0.39,trim=1.5cm 0 0 0]{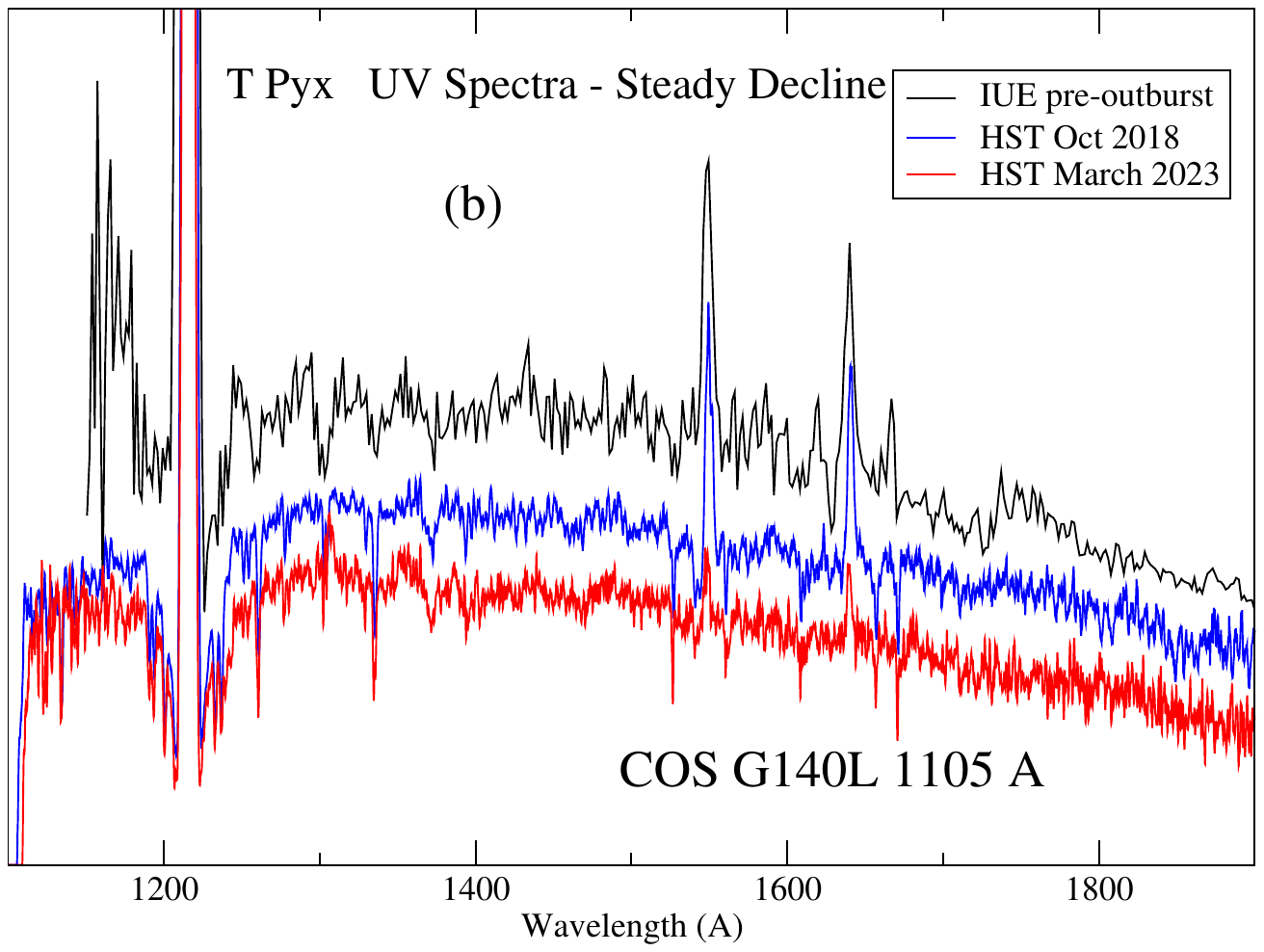}
\caption{
UV spectra of T Pyx showing a steady decline in the continuum flux level.
In the short wavelength region (FUV, left panel - a), the flux drop 
since 2018-2019 is not as large as in the longer wavelength region
(mid-UV, right panel -b). In 2012-2013, the UV flux had reached
its pre-outburst level. In the region where the spectra overlap
($\sim 1150-1200$~\AA ), the short wavelength spectra (a) do not match
the long wavelength spectra (b) as they systematically appear to have 
a higher flux.  
The spectra are presented here before dereddening.  
\label{uvspecdecline} 
}
\end{figure}

We expected the UV decline would have reached a plateau by 2023, 
mimicking the post-outburst IUE Data. However, since Oct 2018 
the UV flux has further dropped by $\sim$20\% and is now about 40\%
below its IUE pre-outburst level - see Fig.\ref{uvspecdecline}.  
%
%
Even the C\,{\sc iv} (1550) and He\,{\sc ii} (1640) emission lines,  
which were prominent in the IUE pre-outburst and HST post-outburst spectra,    
are now much reduced: the intensity of the 
C\,{\sc iv} line in 2023 is $\sim$1/6 of what it 
was in 2018, and that of the He{\sc ii} line is $\sim$1/3.  

Note that the COS G130M 1055 spectra (Fig.\ref{uvspecdecline}a) 
are extremely noisy below $\sim$1090~\AA\ (as seen already in 
Fig.\ref{fuvslines}).
The 1150-1200~\AA\ region of the 2023 COS G130M (1055) 
(red spectrum in Fig.\ref{uvspecdecline}a) 
doesn't match the 2023 COS G140L 1105 spectra 
(also red in Fig.\ref{uvspecdecline}b). 
A similar discrepancy in fluxes was apparent in the October 2015 COS data
of T Pyx \citep{god18} 
between the two configurations (G140L/1105 vs. G130M/1055), 
and, while some of the discrepancy could be attributed to orbital 
modulation, it is mainly due to calibration errors (edges of the detectors). 
The two Si\,{\sc ii} lines (1190.4 \& 1193.3 \AA ) are clearly seen in the 
G130M (1055) spectra (and even in the IUE spectrum; 
right edge of panel (a) of Fig.\ref{uvspecdecline}) 
but they are absent in the G140M (1105) spectra 
(blue and red spectra; left edge of panel (b) in Fig.\ref{uvspecdecline}). 

\clearpage

\begin{figure} 
\plotone{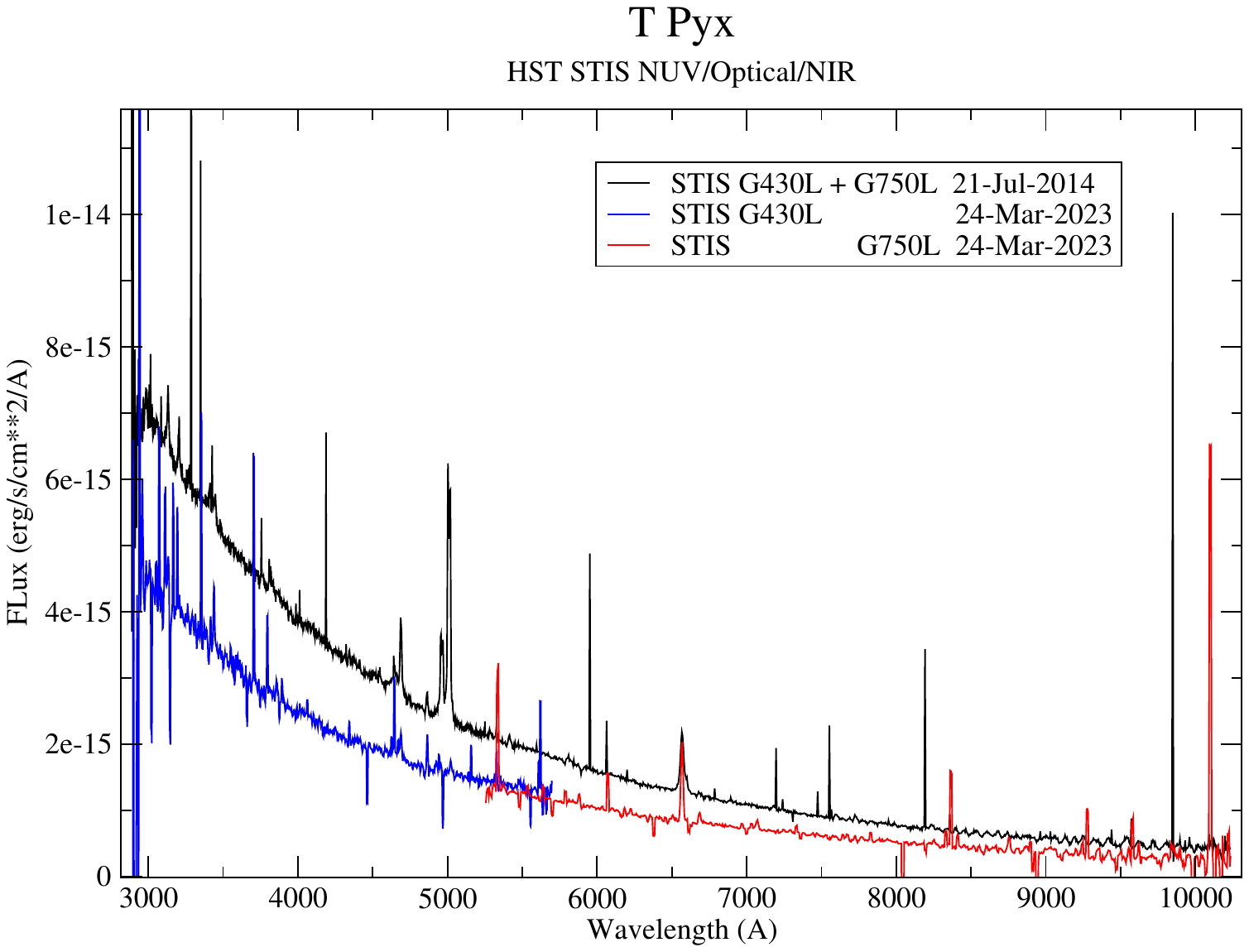} 
\caption{ 
Comparison of the HST optical-NIR spectrum from 2023 
(in blue and red) 
to the same spectrum obtained almost a decade earlier in 2014. 
The spectra have not been corrected for extinction. 
\label{opt23}
}
\end{figure} 

As T Pyxidis erupted in 2011, it became the target of several observing campaigns 
and many optical spectra were obtained with HST/STIS.    
The latest HST optical spectra of T Pyx collected before our current HST 2023 observation 
are from July 2014: OCIQ020 (STIS G430L) and OCIQ030 (STIS G750L),   
made of 16 exposures each (see Table 2).    
All the optical STIS spectra following the outburst and through July 2014 reveal the presence of nebular 
emission lines.  
We extracted the 32 1D spectra from the July 2014 STIS data (as listed in
Table 2) and co-added the 16 exposures each for the OCIQ020 and OCIQ030 sets  
by weighting them by exposure time. We then combined the OCIQ020 and 
OCIQ030 spectra together.  
A comparison with the 2014 optical STIS spectra (Fig.\ref{opt23}) reveals that the optical 
continuum flux level (near $\sim$4,500~\AA ) has now dropped by $\sim 38$\% and the nebular emission
lines have almost all completely disappeared.  
In the very long wavelength range (near $\sim$8,000~\AA ) corresponding to the NIR the
continuum flux level has dropped by $\sim$35\%. 
Note that the 2014 and 2023 spectra displayed in Fig.\ref{opt23} were  
generated by co-adding the individual exposures weighted by the exposure 
time for each of the COS configuration G430 and G750L. 
Since the G430 and G750L spectra covers less than 
the binary orbital period, their continuum flux level did not match perfectly. 
This is most apparent in the 2023 spectra which have shorter exposure times 
($\sim 2000$s) than the 2014 spectra (totalling more than $8,000$s, but covering
only 80\% of the binary orbital period due to the timing of the exposures).  
For the 2014 spectrum, the G750L segment has to be scaled down by $\sim$1\%
to match the G140M segment; 
for the 2023 spectrum, the G750 segment has to be scaled up by 7\%  
to match the G140M segment.

\clearpage

\section{{\bf Analysis}} 

\subsection{{\bf Variability}} 

\begin{figure}[b!]  
\epsscale{0.9} 
\plotone{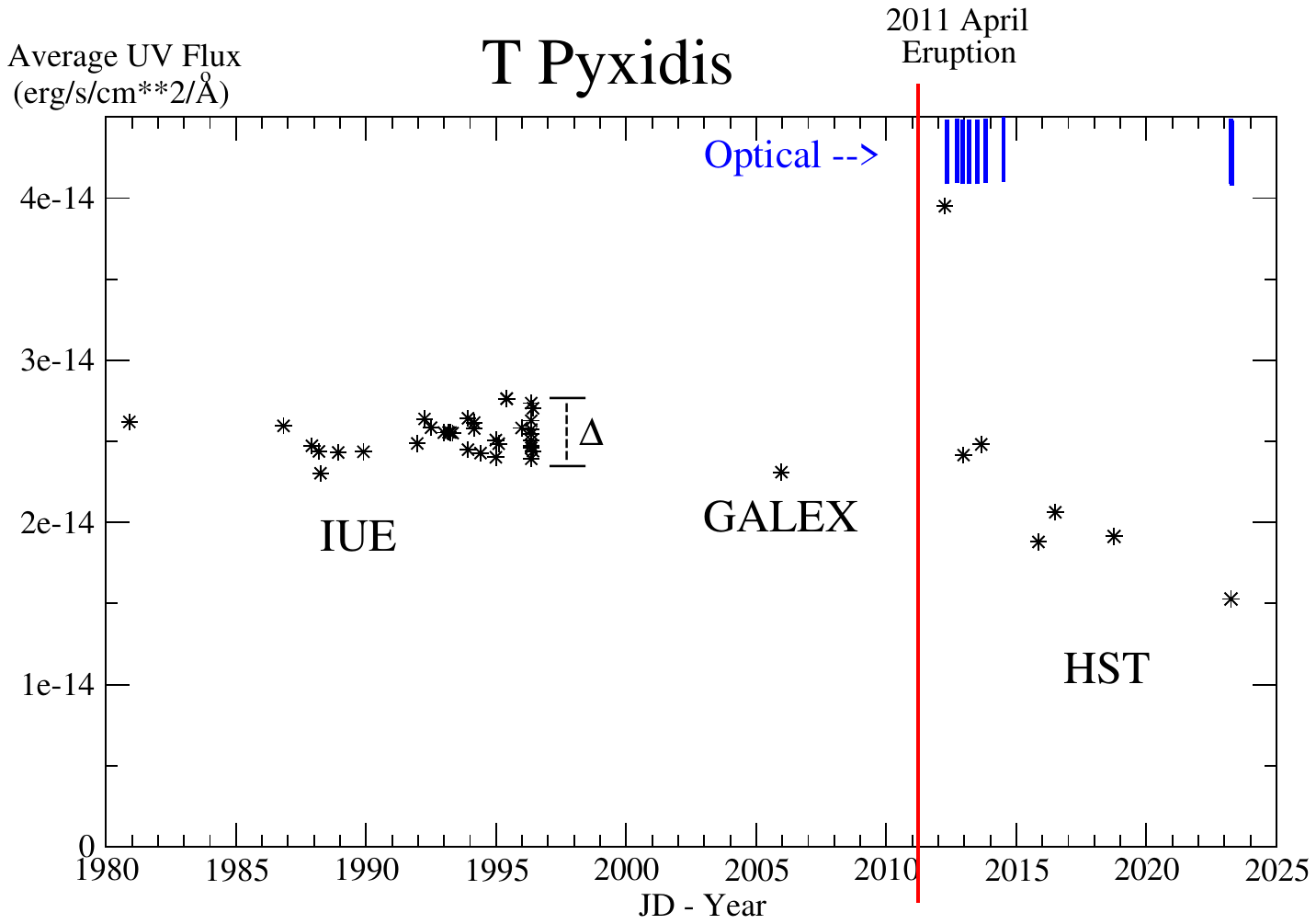} 
\caption{ 
The average UV continuum flux level of T Pyxidis (black stars) 
over the spectral range [1400-1700\AA ] as a function of time 
(the data has not been corrected for extinction). 
The red vertical line represents the 2011 April outburst of the system.
HST STIS optical spectra were also collected after the outburst 
and their timing is indicated using vertical blue lines at the top. 
This graph is an update of the one presented in \citet{god18} and shows
that the UV continuum flux level continues to drop {\it well below its
pre-outburst level.}  
\label{lightcurve}
}
\end{figure}

It has been well documented \citep{sch13} that the (B-band) magnitude
of T Pyx has been steadily decreasing from $B=13.8$
in 1890 to $B=15.7$ just before the 2011 eruption. 
And in recent years, as T Pyx returned to quiescence following the 2011 outburst, 
it has gradually become fainter from 15.8 to 16.1 \citep{waa23}.   
In order to check the behavior of T Pyx in the UV, we generated a UV 
light curve of the system \citep{god18} using archival UV spectra from IUE
(from 1980 to 1996), Galex \citep[one spectrum obtained in 2005, see][]{sch13}, 
and HST STIS \& COS UV spectra (following the 2011 outburst).  
We display in Fig.\ref{lightcurve} an updated UV light curve using
the latest HST spectra (from 2023 and 2018).  
All the data points were obtained by integrating the UV
spectral flux between 1400~\AA\ and 1700~\AA\ (excluding emission and absorption lines).   
The IUE do not reveal a decrease in the continuum flux level 
between 1980 and 1996, and clearly show a modulation 
$\Delta$ of up to 18\% (i.e. $\pm$9\%) in the UV continuum flux level. 
Unfortunately, the IUE single exposures had all a duration of
$\sim$1 to $\sim$2 times the binary orbital period, and a phase resolved UV light
curve could not be generated. 
However, we attribute this modulation to the orbital motion of the binary. 
Since the UV light curve shown in the figure results from an integration
over a large spectral range, the flux error is much reduced and completely
negligible in comparison to the uncertainty due to the orbital 
modulation. 
The HST data points also reveal orbital modulation but to a lesser
extent since not every orbital phase was covered.  
Following the 2018 HST observation, we expected the UV flux to reach a plateau
but the 2023 data point shows a further drop of 20\% compared to the 
2018 data. While we cannot rule out that 
this could be due, in part, to orbital modulation, the UV light curve after
the 2011 outburst definitely exhibits a trend consistent with a steady decline
in sharp contrast with the pre-outburst light curve.

\clearpage 

\begin{figure}[b!]  
\includegraphics[scale=0.35,trim= 0 0 0 0]{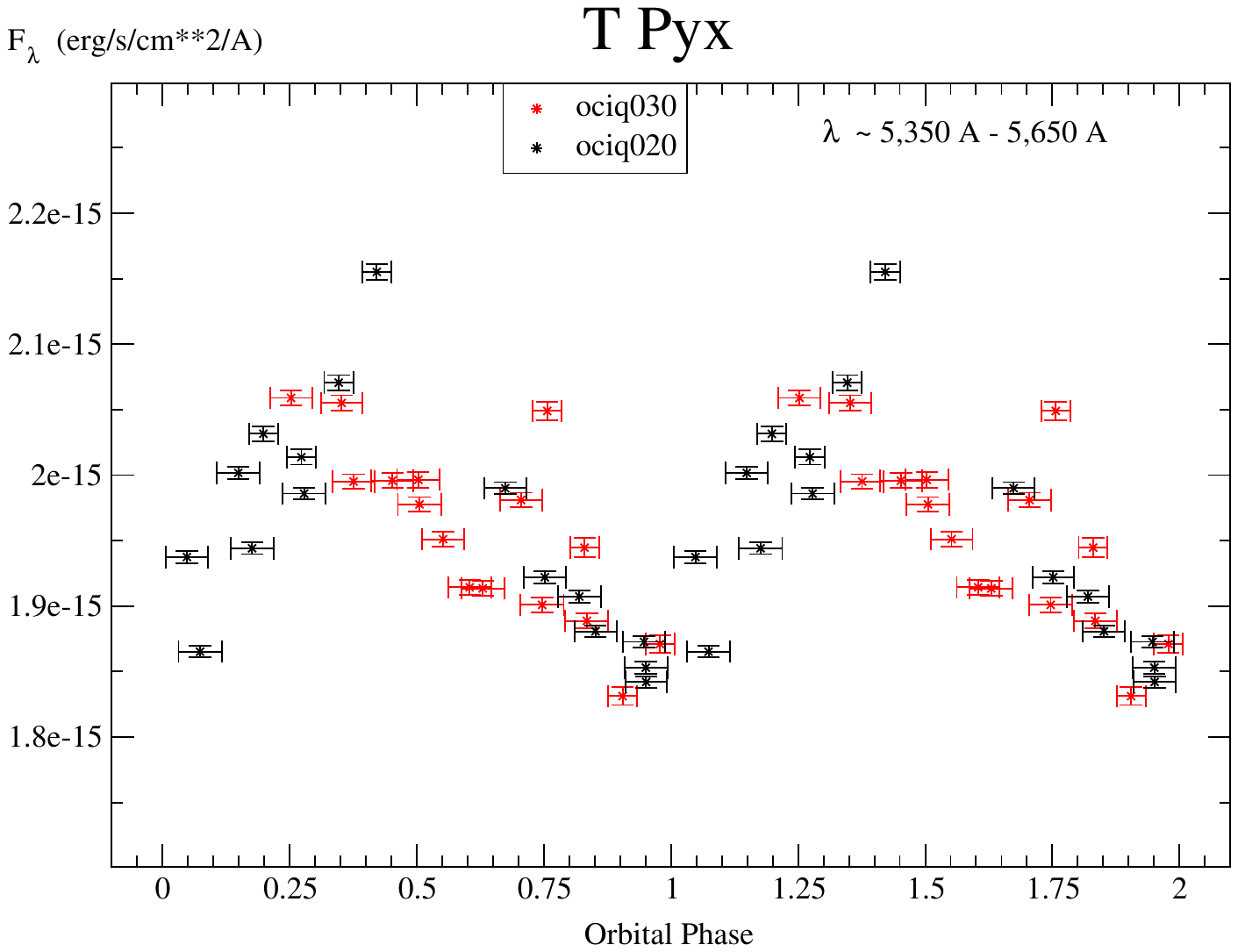}
\includegraphics[scale=0.35,trim= 3.0cm 0 0 0]{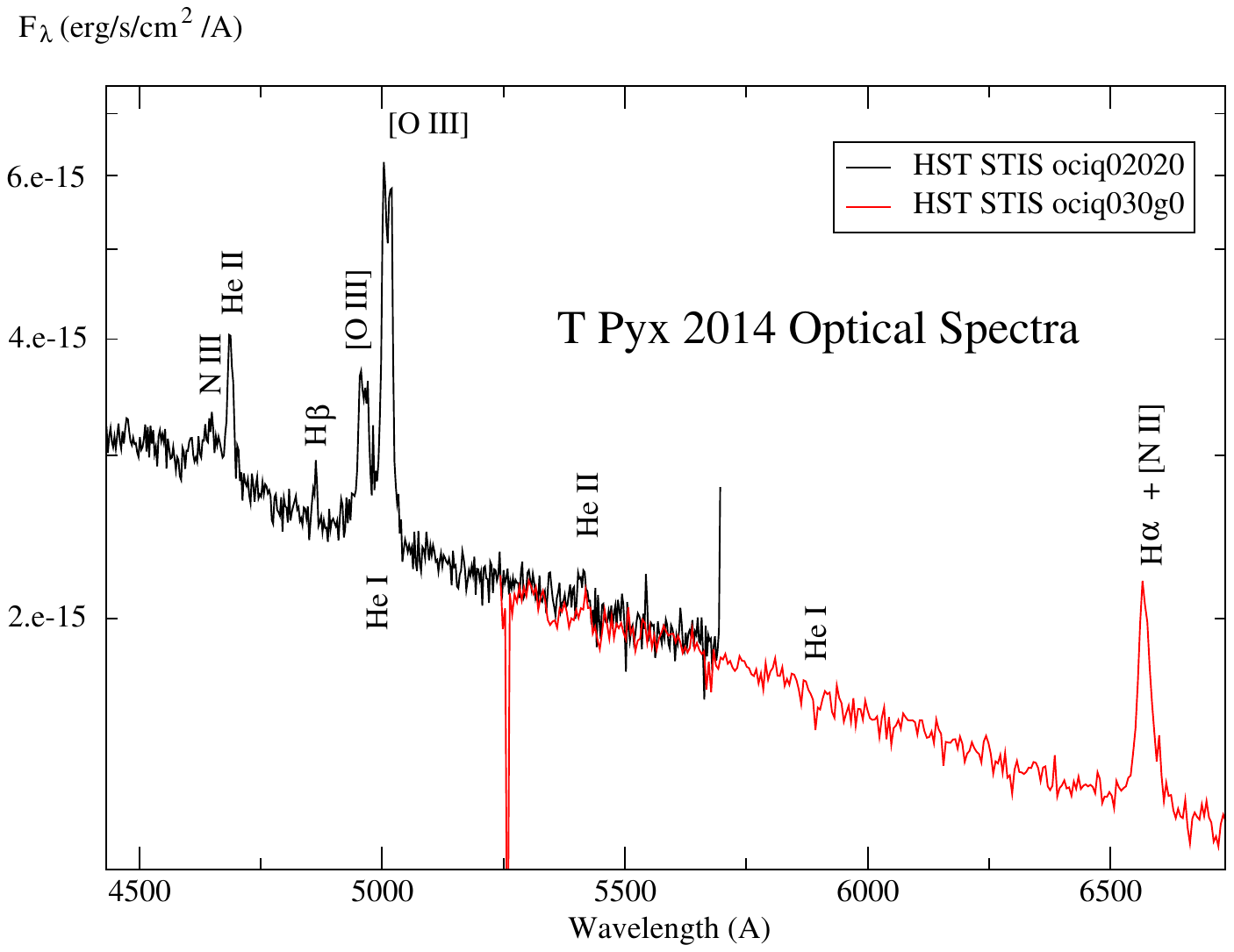}
\vspace{1.cm} 
\caption{
{\bf Left.} 
The {\it chartreuse} (yellow-green color) light curve
of T Pyx folded at the orbital phase revealing 
an orbital modulation.
The chartreuse light curve data were obtained
from HST STIS archival data OCIQ020 (black dots) and OCIQ030 (red dots)  
integrated over the spectral wavelength range $\lambda \sim 5285-5655$~\AA ,  
where the grating G430L (centered at 4300~\AA , for OCIQ020) overlaps with the
grating G750L (centered at 7751~\AA , for OCIQ030). 
The data sets OCIQ020 and OCIQ030 have 16 exposures each (see Table 2). 
{\bf Right.} 
The overlap region of the two STIS datasets (OCIQ020 with G340L) 
and (OCIQ030 with G750L) is shown.   
For clarity, only one exposure for each STIS setting is shown here:
ociq02020 in black and ociq030g0 in red. 
The flux (vertical) axis is displayed on a log scale.     
The data presented in this Figure were not corrected for extinction. 
\label{chartreuse}
}
\end{figure}

As for most novae and recurrent novae, many more HST observations were
carried out during outburst and decline from outburst than during deep quiescence,
and except for our current 2023 STIS optical spectrum,  
all the HST optical spectra were obtained between 2011 and 2014,
while the system still showed nebular emission.  
Among these HST optical spectra we selected the STIS datasets OCIQ020 
(G430L) and OCIQ030 (G750L), each with 16 exposures (see Table 2) from July 2014: 
the emission lines of forbidden transitions forming in the nebular material
are still present (see Fig.\ref{opt23}).  
These two datasets were obtained only two days apart and while they cover
different spectral wavelength regions, they do overlap between 5245~\AA\ and 5690~\AA .    
We therefore integrated the flux of the 32 exposures between 
5285~\AA\ and 5655~\AA\ (excluding emission lines), to compute the average continuum flux level
in that wavelength region, which corresponds the yellow-green color ``{\it chartreuse}''.  
In Fig.\ref{chartreuse} we present the chartreuse light curve folded at the 
orbital phase which clearly reveals the orbital modulation
of the continuum flux level with an amplitude of $\pm 8$\%, similar to the UV data.    
The flux is minimum near phase 0.9, where the L1-stream is hitting the rim of the disk
and indicates that the disk edge might be swollen and partially occulting the disk 
\citep[self-occulting, see][]{pat98}.   
The orbital phase values were computed using the post-outburst ephemerides 
provided by \citet[][equation 2]{pat17} which takes into account the
period change of the system, and taking the mid-value of the 
observation times of each exposure listed in Table 2
(namely we added half the exposure time to the starting time). 
The flux error on the integrated wavelength region is of the
order of $5 \times 10^{-18}$erg/s/cm**2/\AA , 
the error on the orbital phase is taken as (half) the exposure time as listed in 
Table 2 for each exposure (namely 387/2s for the first 4 exposures, 558/2s
for the next 4, etc..).  

\clearpage 

\subsection{\bf{The Spectral Slope}}

\begin{figure}[b!]  
\includegraphics[scale=0.50,trim=-4cm 0 0 0]{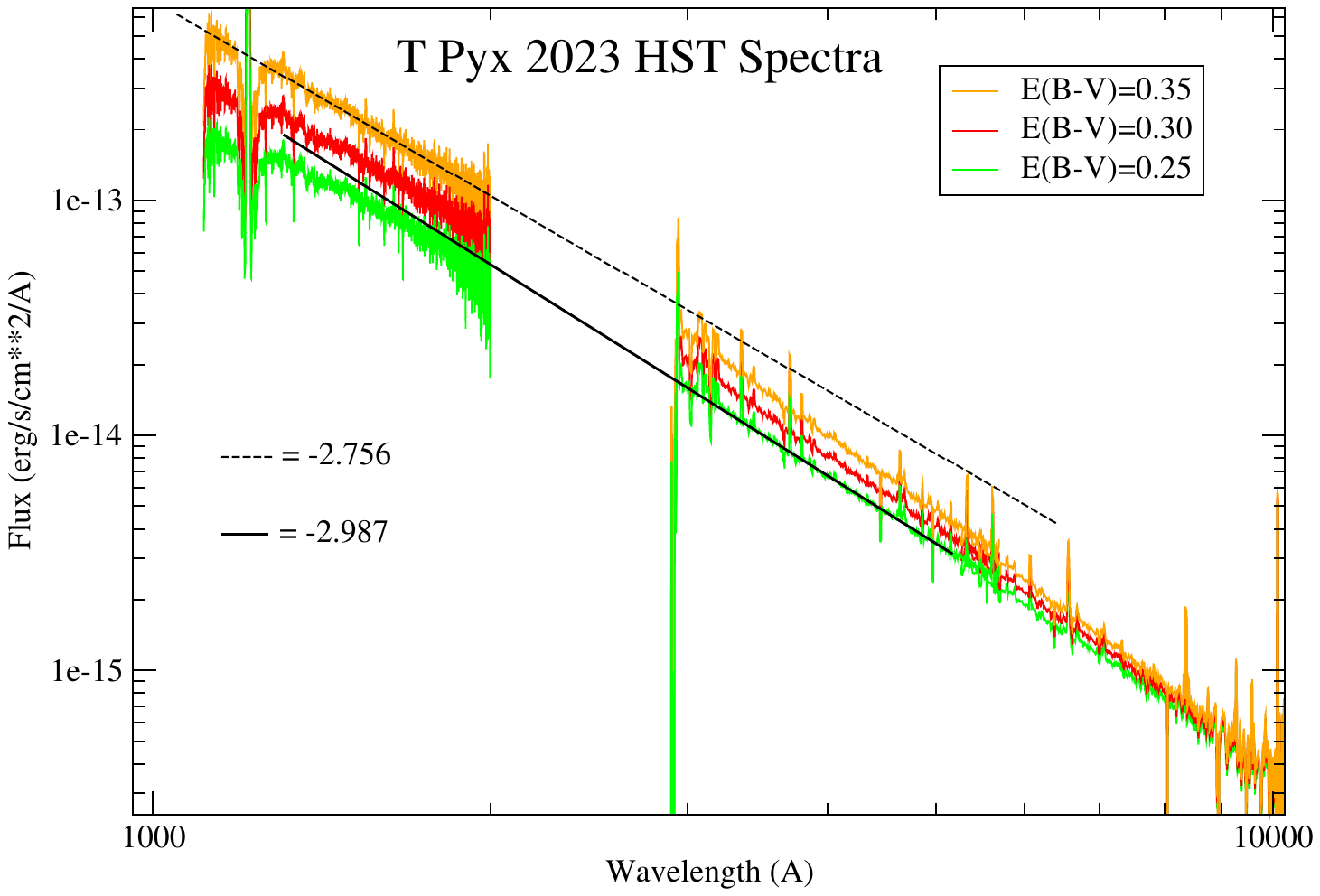} 
\caption{ 
The HST UV-Optical spectrum of T Pyx dereddened for 
$E(B-V)=0.25$ (green), 0.30 (red), and 0.35 (orange), 
on a log-log scale. 
The slope of the continuum flux level is steeper in the
optical than in the UV. 
Even for the larger dereddening, the slope of the continuum
flux level in the UV (-2.76, dashed black line) is not as steep 
as the slope of continuum flux level in the optical,
which has the shallowest slope of (-2.99, solid black line)
for the smallest value of the dereddening considered here.  
\label{diskslope} 
}
\end{figure}

In \citet{god18}, we used archival optical and NUV (IUE and Galex) spectra 
to supplement our HST FUV post-ourburst 
spectra in our accretion disk modeling and 
found that the slope of the continuum flux level is flatter in the 
optical than in the UV. However, these pre-outbursts archival optical and NUV data were 
not obtained concurrently with the same telescope, and the optical data were obtained from
ground-based telescopes and digitally extracted from graphs. 
Therefore, we decided to carry out a new assessment of the slope of the spectrum
using an updated and improved dereddening law and   
using UV and optical data obtained the same day with HST: the 2023 HST STIS and COS spectra.   
In the FUV, instead of using the extinction law of \citet[][as we did in \citet{god18}]{fit07}, 
we used the standard curve of \citet{sav79}, which gives a smaller
correction in the FUV (and therefore a shallower slope in the FUV). 
This is in line with the analysis of T Pyx by \citet{sel13}, based on the work 
of \citet{sas02} who showed that in the FUV the observed extinction
curve is consistent with an extrapolation of the standard extinction curve
of \citet[][further details and discussion on our choice
of the extinction curve were given in  \citet{god20}]{sav79}. 
In Fig.\ref{diskslope} we display the combined 2023 (UV+Optical+NIR) spectrum
of T Pyx dereddened assuming $E(B-V)=0.25$, 0.30, and 0.35 (the values we adopted)
on a log-log scale of the flux $F_\lambda$ (in erg/s/$cm^2$/\AA ) vs wavelength $\lambda$
(in \AA ). 
The steepest (continuum) spectral slope in the UV (at wavelengths longer than
the Ly$\alpha$ region, $\lambda > 1300$~\AA )
is obtained for a dereddening of $E(B-V)=0.35$  
and has value $\alpha=-2.76$, while the flattest optical 
($\lambda \sim 3000-6000$~\AA  ) slope is obtained
for a dereddening of $E(B-V)=0.25$ and gives a slope $\alpha=-2.99$, larger than
the UV.  At longer wavelength (NIR, $\lambda \sim7000-10,000$~\AA ) 
the slope steepens even more ($< -3$).  
Namely, we find that the spectral slope steepens with
increasing wavelength, thereby confirming the findings of 
\citet{gil07}. This finding is valid for the values of 
E(B-V) we (and \citet{gil07}) use when dereddening the spectra for 
the analysis of T Pyx optical and UV data.  
While we found that both the UV and optical continuum flux levels vary  
by about the same amplitude as a function of the orbital phase, 
their slope did not reveal orbital modulation.

\subsection{{\bf Accretion Disk Modeling \& Spectral Analysis}} 

The spectral analysis procedure we follow is extensively described 
in \citet[][]{god20} and only a short overview is given here. 
We use the suite of {\tt FORTRAN} codes {\sc{tlusty}} \& {\sc{sysnpec}} 
\citep{hub17a,hub17b,hub17c} to generate accretion disk spectra 
for a given WD mass $M_{\rm wd}$, mass transfer rate $\dot{M}$, inclination $i$, 
and inner \& outer disk radii $(r_{\rm in},r_{\rm out})$. 
The accretion disk is based on the standard disk model \citep{sha73,pri81},  
it is assumed to be optically thick and has solar composition.   
We generate a grid of disk spectra for 
$M_{\rm wd}=1.0, 1.2, 1.37M_\odot$, $r_{\rm in}=R_{\rm wd}$,  
$10^{-8} M_\odot$/yr$\le \dot{M} \le 10^{-6}M_\odot$/yr (increasing 
or decreasing $\dot{M}$ in steps of $\sim$50\%), and for $i=50^\circ$, $60^\circ$. 
These theoretical spectra extends from 900~\AA\ to 7,500~\AA . 

We first assume a WD mass of $1.37M_\odot$, and it is understood that 
accretion disk model fits with a 
lower WD mass (see further down) will result in a larger mass accretion rate. 
With a secondary mass of $0.13M_\odot$, and an orbital period of 1.8295~hr,  
we obtain a
binary separation of 585,592~km. For such a mass ratio ($log(q)\approx -1.0$), 
the outer radius of the disk is expected to be tidally truncated 
at $r_d \approx 0.5a$ \citep[Fig.3 in][]{goo93}, where $a$ is the binary separation,
while the Roche lobe radius of the WD is about $0.6a$.  
We note that for a mass ratio close to one the tidally truncated disk radius is close to 0.3a 
\citep{pac77}, while for a vanishingly small mass ratio it is close to 0.6a  
\citep{goo93}. 
We compute disk models assuming an outer disk radius of 180,000~km
(90$R_{\rm wd}$) and 360,000~km (180$R_{\rm wd}$) , 
corresponding to about $\sim$0.3a and $\sim$0.6a respectively 
(where we have assumed a 2,000~km radius for T Pyx's WD).
As we lower the WD mass to $1.0 \times M_\odot$ we obtain a binary 
separation much closer to $5 \times 10^5$~km and vary the outer disk 
radius accordingly. 

We first carry out accretion disk spectral fits assuming $M_{\rm wd}=1.37M_\odot$, and for the 
following values of the parameters: $i=50^\circ$ \& $60^\circ$, 
outer disk radius $r_{\rm out}=0.3a$ \& $0.6a$, $E(B-V)=0.25,30,35$ and a Gaia distance of 
2389~pc, 2860~pc, \& 3676~pc (see Table 1 for system parameters). 
For each set of $(i,r_{\rm out},E(B-V),d)$ the spectral fits yields a 
unique value of the mass accretion rate $\dot{M}$.    

\clearpage 

\begin{figure}
\vspace{-1.cm} 
\plotone{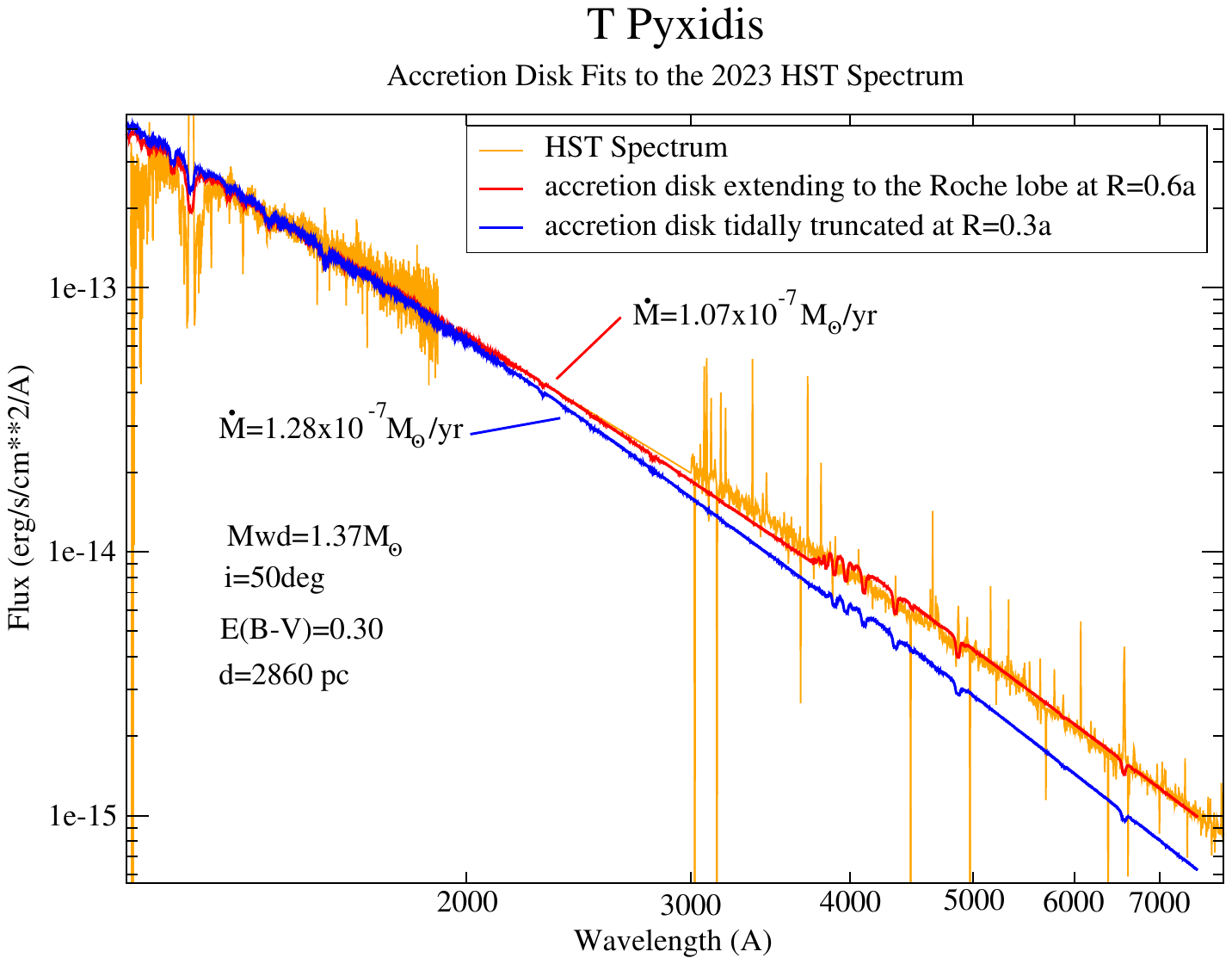} 
\caption{Accretion disk fits to the dereddened HST spectrum of T Pyx 
for a WD mass $M_{\rm wd}=1.37 M_\odot$, inclination $i=50^\circ$, 
and a Gaia distance of 2860~pc. The results are presented on a
log-log scale of the flux vs. wavelength.  
Assuming E(B-V)=0.30, the HST spectrum (orange line) is fitted for a mass transfer
rate $\dot{M}=1.07 \times 10^{-7}M_\odot$/yr when the disk extends to $0.6a$ 
(red line). Assuming a disk truncated at $R=0.3a$ (blue line) increases $\dot{M}$ to 
$1.28 \times 10^{-7} M_\odot$/yr and   produces a stepper slope.
\label{diskfit}
}
\end{figure}

In Fig.\ref{diskfit} we display two of the accretion disk spectral fits we 
ran assuming an inclination of $50^\circ$, extinction $E(B-V)=0.30$, and
the Gaia distance of 2860~pc. In one model the disk was truncated at 
$0.6a$ and in the second model it was truncated at $0.3a$. 
The resulting mass accretion rate is $\dot{M}=1.07 \times 10^{-7}M_\odot$/yr
for the larger disk, versus $1.28 \times 10^{-7}M_\odot$ for the smaller disk. 
Since most of the flux is emitted in the UV range, we mainly fit the 
UV region, and it appears immediately that the smaller disk doesn't
fit the optical as it is too blue. On the other hand, the larger disk 
displays a small Balmer jump which is not seen in the observed spectrum
(we will extend on this issue in the next section).

While we are aware that it is likely that the disk radius is about 0.5a, we did 
ran models for both the 0.3a and 0.6a values. 
For all the values of the parameters considered here 
(and including their uncertainty),  
the results can be summarized as follow: we obtain a mass accretion rate of 
$$                
\dot{M} = 1.38_{-0.87}^{+1.17}\times 10^{-7}M_\odot /{\rm yr} 
$$                
for 
$$i=55^\circ \pm 5^\circ,~~ \frac{r_{\rm out}}{a} =0.45\pm0.15,~~ 
E(B-V)=0.30\pm0.05,~~ {\rm and}~ d=2860^{+816}_{-471}~{\rm pc},
$$ 
assuming a near-chandrasekhar WD mass of $1.37M_\odot$.  
The error in $\dot{M}$ is due mainly to the uncertainty in the distance 
and reddening, while the uncertainty in the value of the outer disk radius 
contributes less than 10\%  to the (relative) error in $\dot{M}$.   

Next we assume a WD mass $M_{\rm wd}=1.2 M_\odot$ and $M_{\rm wd}=1.0 M_\odot$, 
and obtain similar results with a larger mass accretion rate: 
$$                
\dot{M} = 2.16_{-1.36}^{+1.81}\times 10^{-7}M_\odot /{\rm yr}, ~~~~~{\rm and}~~~~~ 
\dot{M} = 2.94_{-1.85}^{+2.46}\times 10^{-7}M_\odot /{\rm yr},  
$$                
for 
$M_{\rm wd}=1.2 M_\odot$ and $M_{\rm wd}=1.0 M_\odot$, respectively.  

\clearpage

\section{\bf{ Discussion \& Conclusion}}  

The present UV-optical analysis shows that the mass accretion in T Pyx
has been steadily decreasing, and is now of the order of $10^{-7}M_\odot$/yr
(based on the assumed system parameters), about 40\% lower than 
its pre-outburst value assessed from archival IUE spectra. 
The decreased {\it activity} of the system is further supported by 
the weakened C\,{\sc iv} (1550) and He\,{\sc ii} (1640) emission lines
(in the COS G140L spectrum from March 2023),   
which were prominent in the IUE pre-outburst and HST post-outburst spectra.    
The mass accretion rate, however, cannot be determined
accurately since the reddening, distance and WD mass have not been
themselves assessed with a high accuracy. The WD mass is assumed
to be large ($\sim 1 M_\odot$ or even near-Chandrasekhar) on theoretical ground (as explained in \S 1), 
the Gaia parallax has a relatively large error, and while we assume
$E(B-V)=0.30\pm0.05$, some authors have derived an extinction
as large as $E(B-V)=0.5\pm0.1$ \citep{sho11}. In our previous
work we showed how the uncertainty in the system parameters
($M_{\rm wd}$, $E(B-V)$, $d$, and $i$) affects the derived 
mass accretion rate by an order of magnitude:  
$\dot{M} \sim 10^{-7\pm 1}M_\odot$/yr \citep[figure 10 in ][]{god18}.  

Another source of uncertainy is the chemical composition of the accretion disk. 
We assume solar abundances for the accretion disk, but the donor/secondary
could have non-solar abundances affecting the shape and slope of the disk spectrum 
(if highly suprasolar/hydrogen deficient). Here too the problem is that the state of 
the secondary star in T Pyx is unknown. 
Absorption lines of {\it metals} (i.e. Z$>$2) for different temperatures in the disk cannot be detected due to the 
combined action of Keplerian broadening and superposition.  
As a consequence, it is the    
hydrogen content (and more precisely the [H/He] ratio) that dictates 
the general shape of the spectrum \citep{god23}, and only for large
values of [H/He] \citep[as observed for the secondary of QZ Ser with a 90\% hydrogen deficit;][]{har18}
is the shape of the spectrum noticeably affected.
Hence, our solar composition results are valid as long as the actual metalicity of the accretion
disk (and therefore secondary donor star) doesn't depart too much from solar.
This is a sound assumption, since even evolved-donor CV systems, with high N and 
low C abundances, are not all strongly hydrogen deficient \citep[unlike QZ Ser, AE Aqr, 
DX And, and EY Cyt;][]{tho02,har18}. 
This is fortunate, since generating accretion disk models from scratch
as a function of chemical abundances is prohibitively CPU-expensive 
(as we already vary the disk input parameters such $M_{\rm wd}$, 
$\dot{M}$, $i$, and the disk radius) and the version of TLUSTY
we are using is not well suited to generate helium dominated spectra. 

In comparison to the above uncertainty in the system parameters, the 
systematic error in the modeling of the disk is rather negligible. 
Whether we chose a disk radius of $0.3a$ or $0.6a$ didn't affect 
the results quantitatively as much as it did qualitatively.   
For a mass transfer rate of 
$\sim 10^{-7} M_\odot$/yr, the temperature in the disk at $r=0.3a$ is 
25,000~K and drops to 15,000~K at $r=0.6a$. As a consequence,
the Balmer Jump is much reduced for an outer disk radius $r_{\rm out}=0.6a$,
and is absent for $r_{\rm out}=0.3a$ because of the higher temperature.
The slope of the continuum flux level of a $r_{\rm out}=0.3a$ accretion disk
is steeper (bluer) than that of a $r_{\rm out}=0.6a$ accretion disk,
but since the smaller disk has a smaller emitting
surface, its flux (for the same $\dot{M}$) is lower than that of the larger disk. 
As a consequence, the smaller disk requires a larger $\dot{M}$ 
(than the larger disk) when fitting the observed spectrum (since the
disk models are scaled to the distance).    
When fitting the HST spectrum, the small disk models were too steep in the 
optical, while those with a large outer radius exhibited a small Balmer jump 
that was not observed. These {\it anomalies} are, however, a well-known
problem in the modeling of CV WDs accreting at a high rate such as novalikes in high
state and dwarf novae in outburst 
\citep[e.g.][]{wad84,wad88,lon91,lad91,lin07,pue07,ham07,god17,gil24}. 
Many suggestions have been advanced to explain the discrepancy and address
the problem, such as modifying the disk radial temperature profile \citep{lon94,pue07,lin10}, 
increasing the inner radius of the inner disk \citep{lin05,god17}. Some suggested irradiation
of the disk \citep{kro07}, others suggested emission from disk winds \citep{mat15}. 
More recently large scale magnetic fields radially transporting angular momentum
and energy have been invoked \citep{nix19}, as well as a disk model where the 
energy dissipation occurs as a function of height ($z$) resulting in 
emission from optically thin regions \citep{hub21}.  
The problem is still a matter of debate and could actually be a combination 
of several of the scenarios suggested here together \citep{zsi24}. 
  
Another source of uncertainty in disk modeling has been the orbital modulation
of the continuum flux level observed both in the UV and optical with a relative
amplitude of 8-9\%. This has been, so far, attributed to the geometry of the
system where the disk likely self-eclipses due to its higher vertical extent
where it is hit by the L1-stream  \citep{pat98,pat17}.    

We note that \citet{kni22} suggest that the unusually high mass accretion rate in T Pyx
could be the result of triple binary evolution. In that scenario,
due to the disturbing effect of a distant companion (the tertiary), 
the inner binary (WD+secondary) orbit can be significantly eccentric,
triggering mass transfer close to periastron passage 
\citep[e.g.][]{sep07,sep09,sep10}. This periodic gas stripping can drive the
secondary out of thermal equilibrium and intense bursts of mass
accretion onto the WD can then kick-start an irradiation-induced 
wind-driven mass-transfer phase \citep{kni00}. 
\citet{kni22} conclude that 
the current high-$\dot{M}$ state of T Pyx is likely associated with 
the high eccentricity of the (inner) binary orbit in the triple system. 
In that case, the light curve variability of T Pyx might not be due
{\it only} to the geometry of the rotating binary system, instead it would
be affected by the periodic mass transfer/stripping near periastron
(periodic increased in $\dot{M}$) and its accretion onto the WD. 
The non-zero eccentricity of the binary orbit is consistent with the possibility of 
the asynchronous rotation raised by \citet{pat17}. 

Though the mass accretion rate we derived of $\sim 10^{-7}M_\odot$/yr cannot be
firmly confirmed due to all the uncertainties cited above, it is consistent
with previous estimates \citep[e.g.][]{pat17}. 
In order for our modeling to agree with 
\citet{sha18}'s accretion rate, we would have to assume a much smaller reddening 
and distance, which would be inconsistent with the Gaia distance and the lower limit
for the reddening. 

One might argue that the relatively high mass accretion rate we obtain ($> 10^{-7} M_\odot$) must result 
in steady nuclear burning on the WD surface and/or affecting the WD structure
(inflating the radius of its outer envelop). 
However, the super-soft X-ray emission in T Pyx began to turn off six months after the
outburst \citep{cho14}, an indication that the nuclear burning on the WD
surface stopped.    
Furthermore, the COS FUV spectral slope 
is consistent with that of an accretion disk and does not accomodate for any
significant contribution from a hot WD component. 
If we try to fit the FUV slope with a single temperature component,
it is consistent with a temperature of the order of 30-40,000~K only. 
This implies that the flux we observed is likely due almost entirely to the accretion disk.      
Interestingly enough, the optical light curve of T Pyx (attributed to the  
L1-inflated disk rim eclipsing the heated secondary) is very similar to the
light curve of one of the prototype supersoft X-ray sources CAL 87 
\citep{mey97}, as already pointed out by \citet{pat98}.   

In spite of all these uncertainties, it is uncontrovertible that the 
UV flux, and therefore the mass accretion rate, is now below its pre-outburst 
level by about 40\%.   
This large decrease in $\dot{M}$ in the $\sim$decade after the 
2011 outburst is in sharp contrast with the rather constant pre-outburst 
UV flux from the IUE spectra.
No such data were collected after the previous outburst for comparison, as  
the earliest IUE spectrum was obtained in 1980, 13$\frac{1}{2}$ years after 
the Dec 66-Jan 67 outburst. 
However, all the IUE spectra obtained through
the 90's have the same continuum flux level as the 1980 IUE spectrum
and show no drop in flux (except for orbital variation). 

For comparison, \citet{sch13} showed that since its outburst in 1890, 
the mass accretion rate in T Pyx has been declining by a factor of 
5.7 in 122~yr, while the UV 40\% decrease in 12~years translates into 
a decline by a factor of $\sim$165 in $\sim$120~yr, 
which is a factor of $\sim 29$ faster. 
Even in the optical, the AAVSO data show a decline of about 1~mag (v) 
since its pre 2011 outburst value to present day, which 
is almost twice as fast as in the 1890-2011 light curve data.  
This could be an indication that the self-sustained feedback loop between the WD
and secondary might is shutting off at an {\it accelerating} rate after the last outburst, 
in agreement with the hibernation theory.   

Observations in the next 5-10 years will be able to assess  
whether $\dot{M}$ continues to drop at such a high rate, 
or whether this is just part of a phase related to 
the decline from the 2011 outburst. 
For example, the UV flux level could reach a plateau within the next coming years, 
based on the post 1966-67 outburst IUE data showing a constant flux level 
through the 80s and 90s, and assuming the behavior of T Pyx is to be the same. 
In that case the drop in $\dot{M}$ is much more pronounced in the decade
following each outburst (and possibly due to the outburst itself).  
Otherwise, if the UV flux continues to drop at the same rate, mass transfer 
will likely completely shut off within a few hundred years and T Pyx 
will enter a hibernation state.  
Our analysis comes to further confirm that T Pyxidis is now in a 
short-lived peculiar phase of its evolution.    

\clearpage  

\begin{acknowledgements}  
Support for this research was provided by NASA through grant number 
HST-GO-17190.001-A to Villanova University from the Space Telescope Science
Institute, which is operated by AURA, Inc., under NASA contract
NAS 5-26555.  
JLS acknowledges support from HST-GO-13400 and NSF AST-1816100. 
We wish to thank the members of the AAVSO who monitored T Pyx in the months
preceeding and up to the HST visit to ensure the safety of the COS instrument
in case of a fast rise to outburst due to an unexpected nova eruption.    
\end{acknowledgements}  

\software{
{\tt{IRAF}} \citep[NOAO PC-IRAF Revision 2.12.2-EXPORT SUN;][]{tod93}, 
{\sc{tlusty}} (v203) {\sc{synspec}} (v48) {\sc{rotin}} (v4) 
\citep{hub17a,hub17b,hub17c}, 
{\tt{PGPLOT}} (v5.2), Cygwin-X (Cygwin v1.7.16),
xmgrace (Grace v2), XV (v3.10) } 
\\ \\

\begin{center} 
{\bf{ ORCID iDs}} 
\end{center} 
Patrick Godon \url{https://orcid.org/0000-0002-4806-5319}  \\ 
Edward M. Sion \url{https://orcid.org/0000-0003-4440-0551} \\  
Robert E. Williams \url{https://orcid.org/0000-0002-3742-8460} \\ 
Matthew J. Darnley \url{https://orcid.org/0000-0003-0156-3377} \\   
Jennifer L. Sokoloski \url{https://orcid.org/0000-0002-8286-8094} \\ 
Stephen S. Lawrence \url{https://orcid.org/0000-0002-7491-7052} 






\begin{thebibliography}{}
\expandafter\ifx\csname natexlab\endcsname\relax\def\natexlab#1{#1}\fi
\providecommand{\url}[1]{\href{#1}{#1}}
\providecommand{\dodoi}[1]{doi:~\href{http://doi.org/#1}{\nolinkurl{#1}}}
\providecommand{\doeprint}[1]{\href{http://ascl.net/#1}{\nolinkurl{http://ascl.net/#1}}}
\providecommand{\doarXiv}[1]{\href{https://arxiv.org/abs/#1}{\nolinkurl{https://arxiv.org/abs/#1}}}

\end{thebibliography}


\begin{thebibliography}{}

\bibitem[Bode \& Evans(2008)]{bod08}
Bode, M.F., \& Evans, A. 2008, {\it Classical Noave} (2nd ed.; Cambridge:
Cambridge University Press) 

\bibitem[Chomiuk et al.(2014)]{cho14} 
Chomiuk, L., Nelson, T., Mukai, K., Sokoloski, J.L., Rupen, M.P. et al. 2014, \apj, 788, 130 

\bibitem[la Dous(1991)]{lad91} 
la Dous, C., 1991, \aap, 252, 100 

\bibitem[La Dous(1994)]{lad94} 
La Dous, C. 1994, Space Science Reviews, Vol.67, p.1 

\bibitem[Darnley et al.(2017)]{dar17}
Darnley, M.H., Hounsell, R., Godon, P., et al. 2017, \apj, 849, 96 

\bibitem[Duerbeck \& Seitter(1979)]{due79}
Duerbeck, H.W., \& Seitter, W.C. 1979, The Messenger, vol.17, p.1 

\bibitem[Fitzpatrick \& Massa(2007)]{fit07} 
Fitzpatrick, E.L, \& Massa, D. 2007, \apj, 663, 320 

\bibitem[De Gennaro et al.(2014)]{deg14}
De Gennaro, A., Shore, S.N., Schwartz, G.J., Mason, E., Starrfield, S. et al. 
\aap, 562, 28 

\bibitem[Gilmozzi \& Selvelli(2007)]{gil07}
Gillmozzi, R., \& Selvelli, P. 2007, \aap, 46, 593 

\bibitem[Gilmozzi \& Selvelli(2024)]{gil24} 
Gilmozzi, R., \& Selvelli, P. 2024, \aap, 681, 83 

\bibitem[Godon \& Sion(2023)]{god23}
Godon, P., \& Sion, E.M. 2023, \apj, 950, 139 

\bibitem[Godon et al.(2014)]{god14}
Godon, P., Sion, E.M., \& Starrfield, S., Livio, M., Williams, R.E., et al. 2014, \apjl, 784, L33 

\bibitem[Godon et al.(2017)]{god17} 
Godon, P., Sion, E.M., Balman, S., Blair, W.P. 2017, \apj, 846, 52 

\bibitem[Godon et al.(2018)]{god18}
Godon, P., Sion, E.M., Williams, R.E., \& Starrfield, S. 2018, \apj, 862, 89 

\bibitem[Godon et al.(2020)]{god20}
Godon, P., Sion, E.M., Szkody, P., \& Blair, W.P. 2020, \mnras, 494, 5244 

\bibitem[Goodman(1993)]{goo93}
Goodman, J. 1993, \apj, 406, 596 
 
\bibitem[Hack et al.(1993)]{hac93} 
Hack, M., Ladous, C., Jordan, S.D., et al. 1993, 
Monograph Series on Nonthermal Phenomena in Stellar Astmospheres - 
NASA SP, Paris: Centre National de la Recherche Scientifique; 
Washington, DC.: NASA, 1993 

\bibitem[Hamilton et al.(2007)]{ham07} 
Hamilton, R.T., Urban, J.A., Sion, EM. et al. 2007, \apj, 667, 1139 

\bibitem[Harrison(2018)]{har18}
Harrison, T.E. 2018, \apj, 861, 102 

\bibitem[Hillman et al.(2020)]{hil20}
Hillman, Y., Shara, M.M., Prialnik, D., Kovetz, A. 2020, Nature Astronomy, vol.4, p.886 

\bibitem[Hillman(2021)]{hil21}
Hillman, Y. 2021, \mnras, 505, 3260 

\bibitem[Hubeny \& Lanz(2017a)]{hub17a}
Hubeny, I., \& Lanz, T. 2017a, A Brief Introductory Guide to TLUSTY
and SYNSPEC, arXiv:1706.01859 

\bibitem[Hubeny \& Lanz(2017b)]{hub17b}
Hubeny, I., \& Lanz, T. 2017b, TLUSTY User's Guide II: Reference Manual, 
arXiv:1706.01935 

\bibitem[Hubeny \& Lanz(2017c)]{hub17c}
Hubeny, I., \& Lanz, T. 2017c, TLUSTY User's Guide III: Operational Manual, 
arXiv:1706.01937 

\bibitem[Hubeny \& Long(2021)]{hub21} 
Hubeny, I., \& Long, K.S. 2021, \mnras, 503, 5534 

\bibitem[Livio \& Pringle(2011)]{liv11} 
Livio, M., \& Pringle, J.E. 2011, \apjl, 740, L18  

\bibitem[Izzo et al.(2024)]{izz24} 
Izzo, L., Pasquini, L., Aydi, E., Della Valle, M., Gilmozzi, R. et al. 2024, \aap, 686 72 
 
\bibitem[Knigge(2019)]{kni19}
Knigge, C. 2019, private communication 

\bibitem[Knigge et al.(2000)]{kni00}
Knigge, C., King, A.R., Patterson, J. 2000, \aap, 364, L75 

\bibitem[Knigge et al.(2022)]{kni22}
Knigge, C., Toonen, S., Boekholt, T.C.N. 2022, \mnras, 514, 1895 

\bibitem[Kramida et al.(2023)]{kra23}
Kramida, A., Ralchenko, Yu, Reader, J., and NIST Team (2023).
{\it NIST Atomic Spectra Database} (ver.5.11), [Online]. 
Available: https://physics.nist.gov/asd 
National Institute of Standards and Technology, Gaithesburg, MD.
DOI: https//doi/org/10.18434/T4W30F  

\bibitem[Kromer et al.(2007)]{kro07} 
Kromer, M., Nagel, T., Werner, K. 2007, \aap, 475, 301  

\bibitem[Linnell et al.(2005)]{lin05} 
Linnell, A.P., Szkody, P., G\"ansicke, B.T., Long, K.S., Sion, E.M., et al. 2005,
\apj, 624, 923 

\bibitem[Linnell et al.(2007)]{lin07} 
Linnell, A.P., Godon, P., Hubeny, I., Sion, E.M., Szkody, P., 2007,
\apj, 662, 1204  

\bibitem[Linnell et al.(2010)]{lin10} 
Linnell, A.P., Godon, P., Hubeny, I., Sion, E.M., Szkody, P., 
2010, \apj, 719, 271   

\bibitem[Long et al.(1991)]{lon91} 
Long, K.S., Blair, W.P., Davidsen, A.F., Bowers, C.W., Van Dyke Dison W., 
et al. 1991, \apjl, 381, L25  

\bibitem[Long et al.(1994)]{lon94} 
Long, K.S., Wade, R.A., Blair, W.P., Davidsen, A.F., Hubeny, I. 1994, 
\apj, 426, 704 

\bibitem[Matthews et al.(2015)]{mat15}  
Matthews, J.H., Knigge, C., Long, K.S., Sim, S.A., Higginbottom, N. 
2015, \mnras, 450, 3331 

\bibitem[Meyer-Hofmeister et al.(1997)]{mey97}
Meyer-Hofmeister, E., Schandl, S., \& Meyer, F. 1997, \aap, 321, 245 

\bibitem[Nixon \& Pringle(2019)]{nix19} 
Nixon, C.J., \& Pringle, J.E. 2019, \aap, 628, A121 

\bibitem[Paczy\'nski(1965)]{pac65}
Paczy\'nski, B. 1965, AcA, 15, 197 

\bibitem[Paczy\'nski(1977)]{pac77}
Paczy\'nski, B. 1977, \apj, 216, 822 
 
\bibitem[Patterson(1984)]{pat84}
Patterson, J. 1984, \apjs, 54, 443 

\bibitem[Patterson et al.(1998)]{pat98}
Patterson, J., Kemp, J., Shambrook, A., et al. 1998, \pasp, 110, 380 
 
\bibitem[Patterson et al.(2017)]{pat17} 
Patterson, J., Oksanen, A., Kemp, J. et al. 2017, \mnras, 466, 581  

\bibitem[Pringle(1981)]{pri81}
Pringle, J.E. 1981, ARA\&A, 19, 137 
 
\bibitem[Puebla et al.(2007)]{pue07} 
Puebla, R.E., Diaz, M.P., Hubeny, I. 2007, \apj, 134, 1923 
 
\bibitem[Sasseen et al.(2002)]{sas02}
Sasseen, T.P, Hurwitz, M., Dixon, W.V., \& Airieau, S. 2002, \apj, 566, 267 

\bibitem[Savage \& Mathis(1979)]{sav79}
Savage, B.D., \& Mathis, J.S. 1979, \araa, 17, 73 

\bibitem[Schatzman(1949)]{sch49} 
Schatzman, E. 1949, Annales d'Astrophysique, 12, 281 

\bibitem[Selvelli et al.(2008)]{sel08} 
Selvelli, P., Cassatella, A., Gilmozzi, R., \& Gonzalez-Riestra, R. 2008, \aap, 492, 787  

\bibitem[Selvelli \& Gilmozzi(2013)]{sel13}
Selvelli, P., \& Gilmozzi, R. 2013, \aap, 560, 49 

\bibitem[Sembach et al.(2001)]{sem01}
Sembach, K.R., Howk, J.C., Savage, B.D., Shull, J.M., \& Oegerle, W.R. 2001, \apj, 561, 573 

\bibitem[Schaefer(2018)]{sch18}
Schaefer, B.E. 2018, \mnras, 481, 3033 

\bibitem[Schaefer et al.(2013)]{sch13}
Schaefer, B.E., Landolt, A.U., Linnolt, M. et al. 2013, \apj, 773, 55 

\bibitem[Schaefer et al.(2010)]{sch10} 
Schaefer, B.E., Pagnotta, A., \& Shara, M. 2010, \apj, 708, 381  

\bibitem[Sepinsky et al.(2007)]{sep07} 
Sepinsky, J.F., Willems, B., Kalogera, V., \& Rasio, F.A., 2007, \apj, 667, 1170

\bibitem[Sepinsky et al.(2009)]{sep09} 
Sepinsky, J.F., Willems, B., Kalogera, V., \& Rasio, F.A., 2009, \apj, 702, 1387

\bibitem[Sepinsky et al.(2010)]{sep10} 
Sepinsky, J.F., Willems, B., Kalogera, V., \& Rasio, F.A., 2010, \apj, 724, 546

\bibitem[Shakura \& Sunyaev(1973)]{sha73}
Shakura, N.I., \& Sunyaev, R.A. 1973, A\&A, 24, 337 

\bibitem[Shara et al.(1986)]{shara86} 
Shara, M.M., Livio, M., Moffat, A.R.J., Orio, M. 1986, ApJ, 311, 163 

\bibitem[Shara et al.(2018)]{sha18} 
Shara, M.M., Prialnik, D., Hillman, Y., \& Kovetz, A. 2018, \apj, 860, 110 

\bibitem[Shore et al.(2011)]{sho11}
Shore, S.N., Augusteijn, T., Ederoclite, A., \& Uthas, H. 2011, \aap, 533, L8 

\bibitem[Sokoloski et al.(2013)]{sok13}
Sokoloski, J., Crotts, A.P.S., Lawrence, S., \& Uthas, H. 2013, \apjl, 770, L33 
 
\bibitem[Starrfield et al.(2020)]{sta20}
Starrfield, S., Bose, M., Iliadis, C. et al. 2020, \apj, 895, 70 
 
\bibitem[Starrfield et al.(1985)]{sta85}
Starrfield, S., Sparks, W.M., \& Truran, J.W. 1985, \apj, 291, 136    

\bibitem[Starrfield et al.(1972)]{sta72}
Starrfield, S., Truran, J.W., Sparks, W.M., \& Kutter, G.S. 1972, \apj, 176, 169  

\bibitem[Thorstensen et al.(2002)]{tho02}
Thorstensen, J.R., Fenton, W.H., Patterson, J., et al. 2002, \pasp, 114, 1117 

\bibitem[Tody(1993)]{tod93} 
Tody, D. 1993, in ASP Conf. Ser. 52, Astronomical Data Analysis
Software and Systems II, ed. R.J. Hanisch, R.J.B. Brissenden, 
\& J. Barnes (San Fransisco, CA;ASP), 173 

\bibitem[Tofflemire et al.(2013)]{tof13}
Tofflemire, B.M., Orio, M., Page, K.L., et al. 2013, \apj, 779, 22 

\bibitem[Uthas et al.(2010)]{uth10}
Uthas, H., Knigge, C., \& Steeghs, D. 2010, \mnras, 409, 237 

\bibitem[Waagen et al.(2023)]{waa23} 
Waagen, E., O'Meara, S., Poxon, M., Cyanmon, C. 2023, private communication 

\bibitem[Wade(1984)]{wad84}
Wade, R.A. 1984, \mnras, 208, 381 

\bibitem[Wade(1988)]{wad88}
Wade, R.A. 1988, \apj, 335, 394  

\bibitem[Webbink et al.(1987)]{web87} 
Webbink, R.F., Livio, M., Truran, J.W., \& Orio, M. 1987, \apj, 314, 653  

\bibitem[Whelan \& Iben(1973)]{whe73}
Whelan, J., Iben, I. 1973, \apj, 186, 1007 

\bibitem[Yaron et al.(2005)]{yar05}
Yaron, O., Prialnik, D., Shara, M.M., Kovetz, A. 2005, \apj, 623, 398 

\bibitem[Zsidi et al.(2024)]{zsi24}
Zsidi, G., Nixon, C.J., Naylor, T., \& Pringle, J.E. 2024, \mnras, in press 
(arXiv:2406.03676)  

\end{thebibliography}
\end{document}